%% file: main.tex
\documentclass[sigconf]{acmart}

\AtBeginDocument{%
  }

\setcopyright{acmlicensed}
\copyrightyear{2024}
\acmYear{2024}
\acmDOI{XXXXXXX.XXXXXXX}

\acmISBN{978-1-4503-XXXX-X/18/06}

\usepackage{caption}
\usepackage{xcolor}
   \definecolor{codegreen}{rgb}{0,0.6,0}
   \definecolor{codegray}{rgb}{0.5,0.5,0.5}
   \definecolor{codepurple}{rgb}{0.58,0,0.82}
\usepackage{threeparttable}
\usepackage{booktabs}
\usepackage{tikz}
\usepackage[]{graphicx}
\usepackage{subfig}
\graphicspath{{./pictures/}}
\usepackage[ruled,linesnumbered,noend]{algorithm2e}
\usepackage[noend]{algpseudocode}
\SetKwInput{KwInput}{Input}
\SetKwInput{KwOutput}{Output}

\usepackage{booktabs}
\usepackage{bm}
\usepackage{multirow} 
\usepackage{multicol}
\usepackage{flushend}
\usepackage{makecell}
\usepackage{threeparttable}
\usepackage{footnote}
\usepackage{lipsum}
\usepackage{xcolor}
\usepackage{enumitem}
\usepackage{fancyhdr}
\usepackage{tabularx}
\usepackage{svg}
\usepackage{soul}
\usepackage{tablefootnote}
\usepackage{refcount}
\usepackage{pifont}
\newcommand{\cmark}{\ding{51}}%
\newcommand{\xmark}{\ding{55}}%
\usepackage{placeins}
\usepackage{float}
\usepackage[normalem]{ulem}
\usepackage{listings}
    \lstdefinestyle{mystyle}{
        frame=single,
        float = th,
        floatplacement=tbp,
        commentstyle=\color{codegreen},
        keywordstyle=\color{blue}\bfseries,
        numberstyle=\tiny\color{codegray},
        stringstyle=\color{codepurple},
        basicstyle=\linespread{0.5}\fontsize{6}{10.8}\ttfamily\bfseries,
        breakatwhitespace=false,
        breaklines=false,
        captionpos=b,
        keepspaces=true,
        numbers=none,
        showspaces=false,
        showstringspaces=false,
        showtabs=false,          
        tabsize=2,
        language=C,
        moredelim=**[is][\underbar]{_}{_},
        escapechar=\@,
        emptylines=1,
        numbers=left,
        numberstyle=\tiny\sffamily,
        numbersep=4pt,
        aboveskip=4pt,
        xleftmargin=4pt,
        frame=none,
        escapeinside={(*@}{@*)}
    }
\DeclareMathAlphabet{\mathpzc}{OT1}{pzc}{m}{it}
\newcommand{\tech}{\textsc{NatGVD}}
\newcommand{\example}[1]{\parbox{0.33\textwidth}{\lstinputlisting{example_#1.txt}}}
\DeclareRobustCommand{\hlcyan}[1]{{\sethlcolor{yellow}\hl{#1}}}
\DeclareRobustCommand{\hlred}[1]{{\sethlcolor{red}\hl{#1}}}
\DeclareRobustCommand{\hlgreen}[1]{{\sethlcolor{green}\hl{#1}}}

\DeclareMathAlphabet{\mathbbold}{U}{bbold}{m}{n}
\begin{document}

\title{\tech{}: Natural Adversarial Example Attack towards Graph-based Vulnerability Detection}

\author{Avilash Rath$^{\S}$
, Weiliang Qi$^{\S}$, Youpeng Li, Xinda Wang}
\affiliation{
    \institution{The University of Texas at Dallas}
    \country{}
    \thanks{$^{\S}$The first two authors contributed equally to this work.}
}
\email{{avilash.rath, 
 weiliang.qi, youpeng.li, xinda.wang}@utdallas.edu}

\renewcommand{\shortauthors}{A. Rath, W. Qi, Y. Li, and X. Wang}

\begin{abstract}
\input{0_abstract}
\end{abstract}



\received{20 February 2007}
\received[revised]{12 March 2009}
\received[accepted]{5 June 2009}

\maketitle

\input{1_introduction}
\input{2_preliminaries}
\input{3_problem_formulation}

\input{4_transformations}

\input{5_evaluation}
\input{6_discussion}
\input{7_related_work}

\input{8_conclusion}

\bibliographystyle{ACM-Reference-Format}
\bibliography{reference}

\appendix
\clearpage
\input{appendix_h}
\input{appendix_g}
\input{appendix_a}
\input{appendix_b}
\input{appendix_c}
\input{appendix_d}

\end{document}

%% file: 0_abstract.tex
Graph-based models learn rich code graph structural information and present superior performance 
on various code analysis tasks.
However, the robustness of these models against adversarial example attacks in the context of vulnerability detection remains an open question. This paper proposes \tech{}, a novel attack methodology that generates natural adversarial vulnerable code to circumvent GNN-based and graph-aware transformer-based vulnerability detectors. \tech{} employs a set of code transformations that modify graph structure while preserving code semantics. Instead of injecting dead or unrelated code like previous works, \tech{} considers naturalness requirements: generated examples should not be easily recognized by humans or program analysis tools. With extensive evaluation of \tech{} 
on state-of-the-art vulnerability detection systems, 
the results reveal up to 53.04\% evasion rate across GNN-based detectors and graph-aware transformer-based detectors. We also explore potential defense strategies to enhance the robustness of these systems against \tech{}.

%% file: 1_introduction.tex
\section{Introduction}
The increasing prevalence of software in the modern era has led to 
more systems being open to attack by cybercriminals.
Surges in the presence of 
software vulnerabilities pose a significant risk to industrial enterprises' systems, infrastructures, and data. 
For example, the zero-day Log4Shell vulnerability (CVE-2021-44228) affected a large number of businesses using the popular Apache Log4j logging framework, with average loss of around \$90,000 per security breach~\cite{arctic2022log4shell_retrospective}. Integrating effective and robust vulnerability detectors into software development processes is of vital importance for preventing these costly incidents.
To this end, learning-based vulnerability detection techniques can 
automatically detect software vulnerabilities, enabling prompt defense against potential threats.

Program source code contains rich structural information such as control and data dependencies. Benefiting from this information in code graph representations (e.g., program dependency graph), graph-based deep learning (DL) methods, including graph neural network (GNN)-based models and graph-aware transformer-based models, demonstrate superior effectiveness over non-graph-based models for various code analysis tasks, as shown in Microsoft's CodeXGLUE leaderboard~\cite{lu2021codexglue, codexglue_leaderboard}. 
The success of the state-of-the-art (SOTA) graph-based models~\cite{alon2019code2vec, guo2020graphcodebert, hanif2022vulberta, chakraborty2020deep, nguyen2022regvd, guo2022unixcoder} across various code analysis tasks~\cite{codexglue_leaderboard} highlights the need to focus on understanding the weaknesses that exist in these models. Considering the importance of vulnerability detection (VD) to software security practices, we examine recent representative graph-based VD models to evaluate their robustness to potential adversaries.

One of the major problems with these graph-based VD models is their susceptibility to adversarial example attacks. In GitHub, anyone can suggest code changes to public repositories via pull requests. Since the existence of vulnerability detectors is not secret, malicious actors can continuously attempt to craft vulnerable code in a seemingly benign format to evade detection. Existing studies on attacking DL models for code analysis are effective~\cite{li2022closer,zhang2023black,gao2023discrete,yang2022natural,applis2021assessing,imgrund2023broken,yefet2020adversarial,zhang2020generating}, but have three limitations. 
First, few works focus on attacking software VD systems. Although the models for different code analysis tasks may share some similarities, VD relies on extracting unique vulnerable code patterns, which are rare in normal programs. Therefore, general attack methodologies for code analysis models may not work effectively for VD models. 
Second, some attacks 
focus on one specific DL model (e.g., GCN, BERT)~\cite{li2023black_bagammo,yang2022natural}, failing to evaluate attack generalization capability on varied model architectures.
Third, existing works pay little attention to whether their generated adversarial examples (i.e., crafted code) are natural to humans. Since modern software development usually requires code review~\cite{bacchelli2013expectations}, 
unnatural code patterns (e.g., injection of dead or unrelated code) can be easily identified by humans or program analysis tools. Consequently, a successful adversarial attack needs to not only preserve program semantics but also guarantee naturalness to humans and typical program analyzers.

To generate effective adversarial examples for graph-based VD models, our work addresses the following realistic issues:
(1)~\uline{\textit{Vulner-}\\\textit{able} \textit{code analysis focus}}. Instead of adversarial example attacks against general code analysis, our study specifically focuses on models for vulnerable code detection. 
With such targeted attack and evaluation for VD models, their robustness can be revealed.
(2)~\uline{\textit{Code graph focus}}. Unlike existing works on graph-unaware pre-trained models~\cite{li2022closer, zhang2023black, gao2023discrete, wang2023recode, applis2021assessing, yang2022natural}, extensive evaluation on a series of SOTA graph-based VD models should be conducted for a generalizable attack. (3)~\uline{\textit{Natural adversarial example focus}}. The adversary in most scenarios may face multiple directions of scrutiny: machine verification and human verification. Adversarial code should not be easily recognized by humans or program analysis tools. Otherwise, it may be meaningless in a real-world security context. 

In consideration of these issues, we propose \tech{}, a \textbf{\underline{N}}atural adversarial example \textbf{\underline{AT}}tack for \textbf{\underline{G}}raph-based \textbf{\underline{V}}ulnerability \textbf{\underline{D}}etection models.
In \tech{}, we adopt the technique of semantics-preserving program transformations (SPTs) and design a set of natural transformation rules, including assignment splitting, conversion between \texttt{\small while} and \texttt{\small for} loops, condition negation, condition splitting, condition reordering, and their combinations. 

We conduct comprehensive evaluation on the effectiveness of our proposed attack. 
Our experiments also confirm the efficiency of generating adversarial examples and resilience of \tech{} to common defense techniques (i.e., ensemble models and adversarial training). Additionally, we conduct a user study to verify the naturalness of our transformation rules from the perspective of developers. The results show that participants consistently agree that our generated adversarial code samples look natural to them.


The main contributions of our work are summarized below:
\begin{itemize}[leftmargin=*]
\item To the best of our knowledge, we are the first to propose the natural adversarial example attack towards graph-based deep learning approaches for VD. We will release the source code and data artifacts after acceptance.

\item We design a set of natural semantics-preserving transformations (SPTs) that specifically target graph-based vulnerability detectors and verify these transformations are natural to humans.

\item We evaluate the robustness of \tech{} on SOTA GNN-based and graph-aware transformer-based VD models and show that all of these models are vulnerable to our attacks.

\item We explore the resilience of \tech{} to common adversarial defense techniques and provide insights for defenders.

\end{itemize}



%% file: 2_preliminaries.tex
\vspace{-0.08in}
\section{Preliminaries}\label{sec:preliminaries}

\noindent\textbf{Graph Representation for Program Code.} Unlike images (consisting of pixels) and sequential natural language, programming language contains rich semantic structural information. Existing works have demonstrated the effectiveness of employing graph structures to represent source code programs in code analysis-related tasks~\cite{zhou2019devign,chakraborty2020deep, mirsky2023vulchecker}.
One typical graph representation for code is the abstract syntax tree (AST), a hierarchical tree structure that encodes the syntactic structure of source code. 
Other representations include control flow graphs (CFG), control dependency graphs (CDG), data dependency graphs (DDG), as well as various combinations and customizations of these graphs, e.g., code property graphs (CPGs). CFGs and CDGs capture control flow and dependencies between program elements or statements, representing the order of program execution. DDGs represent the dependency of values from definitions of variables to their uses in statements.
Instead of considering code as a sequence of tokens, these graph representations allow for the capture of rich code syntactic and semantic information, enabling learning of more effective vector representations through subsequent neural networks.

\begin{figure}[htbp]
\centering
\captionsetup{justification=centering}
\includegraphics[width=0.75\linewidth, trim=1cm 0cm 1cm 0cm]
{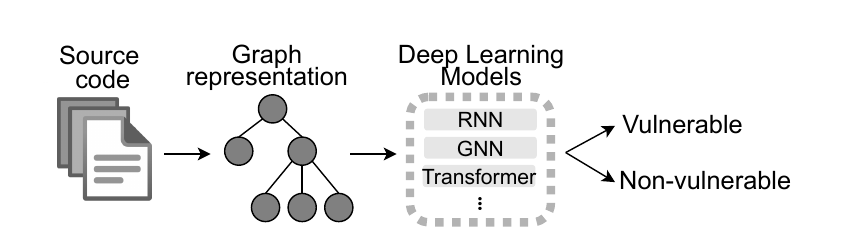}
\caption{Workflow of graph-based DL approach for VD.}
\Description{A graph-based VD workflow. The workflow begins with source code, which is then transformed into a graph representation via a code-to-graph mapping. The graph representation is transformed to vector representation via graph-to-vector mapping. The vector representation is finally fed through a classifier (for example, an RNN, GNN, or Transformer-based model) to generate a prediction. The two types of predictions are vulnerable or non-vulnerable, in this case.}
\label{fig:workflow}
\end{figure}

\noindent\textbf{Graph-based DL Approaches for VD}. Since graphs are a naturally suited way to describe source code, a line of VD methodologies have been proposed to utilize graph representations in their source code-to-prediction processing flows,
as illustrated in~\autoref{fig:workflow}. Known as graph-based VD models, they utilize the graph-based representations of input source code to learn code structural information and 
use this knowledge to predict whether input code is vulnerable. 
Compared to non-graph-based approaches, works that combine code graph representations with various techniques significantly improve VD performance. Examples of such works include ones that adopt code graph representation with sequential Recurrent Neural Networks (RNNs)~\cite{li2018vuldeepecker,li2022sysevr}. Other works attempt to use GNNs to extract semantic structural information from code graph representations ~\cite{zhou2019devign,chakraborty2020deep,nguyen2022regvd,cheng2021deepwukong,wang2020funded,wu2021vulnerability}. Recent DL detection works employ pre-trained transformers. Some of them utilize language models (e.g., CodeBERT) for code graph embedding~\cite{hin2022linevd,he2023bingo}, and others incorporate vulnerable code graph information during model pre-training~\cite{guo2020graphcodebert,guo2022unixcoder,liu2024pre}.
Among them, GNNs and graph-based pre-trained language models~\cite{zhou2019devign,chakraborty2020deep,nguyen2022regvd, guo2022unixcoder} show superior results. Therefore, our work focuses on these models, referring to them as graph-based VD models.

\noindent\textbf{Adversarial Example Attack}.
Adversarial attacks are specially designed manipulations or tricks designed to fool machine learning-based models into making mistakes and producing incorrect results~\cite{madry2018towards}. 
Adversarial example (AE) attacks are a specific kind of adversarial attack which involve crafting or designing a method to produce specific data samples (adversarial examples) that exhibit certain divergences from regular input data for these machine learning models.
In the context of code, adversarial example attacks are usually performed by crafting input code that closely resembles original inputs but causes incorrect behavior by the target model~\cite{yang2022natural}. Types of incorrect model behaviors include misclassification in clone detection or VD tasks, and lowered output quality 
in tasks like bug fix/comment generation. For example, recent work by Li et al.~\cite{li2023black_bagammo} describes an adversarial example attack that targets Function Call Graph (FCG)-based malware detection systems. By inserting never-executed function calls through a ``try-catch trap'' into malware code, FCG-based malware detectors will erroneously classify such malware as benign code.

Other approaches for attacking machine learning-based models in the context of code range from simple to increasingly complex in how they maintain adversarial code consistency. Some earlier approaches~\cite{yefet2020adversarial} used variable renaming and dead code insertion. Adding complexity, a following approach involved the adaptation of the Metropolis-Hastings algorithm~\cite{metropolis1953equation} to 
variable modification in MHM~\cite{zhang2020generating}. Recently, SPTs which ensure code consistency at a more complex level have been employed against a series of models (e.g., GGNN, GNN-FiLM, CodeBERT, code2vec, code2seq, ASTNN, etc.)~\cite{applis2021assessing, pour2021search, rabin2021generalizability, zhang2023challenging}. 
We note that simple SPTs such as variable renaming may not affect code graph structures like CDGs and DDGs, and are therefore less likely to affect graph-based code analysis models. Also, although transformations like dead-code insertions are able to mislead models, they can be easily identified by humans or static/dynamic analysis tools. Therefore, we propose to perform natural transformations with higher complexity to attack recent graph-based VD models.


%% file: 3_problem_formulation.tex

\section{Problem Formulation}\label{sec:formulation}


\subsection{System \& Threat}

In the age of open-source software, we consider adversaries to be malicious users of version control systems like GitHub, GitLab, etc. In these systems, any user can submit a pull request to propose code changes to a public software repository. Repository maintainers review these pull requests using a set of automatic code analysis tools, including graph-based VD models, to decide whether to merge the code. For proprietary software that is not hosted in Git-like systems, 
code submitted by inside developers also needs to go through vulnerability detectors before being used. 

We assume an adversary launches a black-box AE attack towards one of the target graph-based VD systems. To execute the attack, the adversary makes modifications to a host of known vulnerable functions. The vulnerable functions after modification are called adversarial examples. The attacker's goal is to hide their vulnerable code via modification so that it can no longer be classified as vulnerable by the target VD system. By committing such vulnerable code that can bypass the VD models to the benign software repository, the attackers can utilize and exploit these vulnerabilities later.

The adversary knows that the target VD system utilizes a graph-based representation for source code. However, the adversary does not have access to the internal workings or source code of the system. Information about gradients, non-code features, output probabilities, graph generation methods, specific classifier details, parameters, and data granularity is also not available to the attacker. The most significant part of the information available to the attacker is the VD prediction made by the target system; that is, the attacker only knows whether a certain function has been classified as vulnerable or non-vulnerable. This is known as a black-box setting. The defender is assumed to be resistant to attacks with recognizable graph signatures and has sufficient static/dynamic/manual analysis capabilities to detect attacks based on the usage of dead code or injection of unrelated executable code. \tech{} provides an approach for an attacker that works within these restrictions that operates through transformations of 
code.

\subsection{Requirements for Attack Formulation} \label{subsec:requirements}

\tech{}'s attack formulation is aligned with the idea of functional consistency (i.e., the requirement that transformations do not alter the behavior of the modified code). However, there are additional restrictions that must be considered from the attacker's point of view. Transformations made by the attacker must avoid predictable manipulations. Specifically, attacks that involve predictable attack elements may be mitigated using adversarial re-training or signature recognition once the attack methodology has been exposed to the defender~\cite{zhang2009semantic}. Our method avoids being detected via predictability because it focuses on structural changes to code and operates independently of specific vulnerable functions or identifier names. Also, an attacker must consider that adding dead code or extra unrelated code may be detectable using static or dynamic analysis tools. 
Thus, such transformations should be avoided. In Bielik et al.~\cite{bielik2020adversarial}, adversarial robustness is increased by learning relevant parts of an input program and using only those parts as input to the neural network classifier. A skilled adversary would try to apply transformations to relevant parts of an input program, reducing the chance that robust models would ignore the transformations entirely. This can be achieved by using a diverse set of transformations that can be applied to multiple different locations within input programs.

We consider naturalness of code as an important requirement for adversarial vulnerable code produced by an attacker. If code looks unnatural to humans, it will have a high probability of being caught in manual review processes, which is not desired by an attacker. We consider code to be natural if it has a reasonable chance of being produced by an average developer. It should avoid unnatural code structures, variable naming, etc.

Attacks that involve manipulating assignment, call, or control-flow statements (e.g. \texttt{\small{return}}, exit, \texttt{\small{goto}} statements) can potentially affect functional consistency. An attacker must specifically filter the target locations for some transformations to ensure that functional consistency is preserved. Altering conditional statements (e.g. \texttt{\small{if}}, \texttt{\small else}, and loop statements), on the other hand, will not affect functional consistency, provided that order between side effects of sub-expressions is preserved. This kind of transformation can be determined in terms of the sub-expressions of the original code and could be more resistant to defenses that retrain against transformation elements. 
The usage of dead nodes in the code-based graph representation can be avoided by transformations that focus on conditional statements, as these transformations will not create dead nodes. The usage of dead code can also be avoided in this way, since the conditional statements are live code. 
Also, conditional statements are generally associated with multiple source lines of code, which can give them an increased chance of being related to a vulnerable part of an input program. Specifically, the conditional expression(s) in a loop or \texttt{\small if}-\texttt{\small else} statement determines whether the loop/statement body would be executed. A step-by-step description of how these kinds of statements can be involved in the transformation application process can be found in Section \ref{subsec:apply-transform}, preceded by an explanation of transformations used in this work.

\subsection{Attack Formulation}\label{subsec:attack_formulation}

The attack is defined in terms of semantics-preserving transformations (SPTs) being applied to a piece of vulnerable code. Each classification process utilizes a predictive classifier (a GNN model or graph-based transformer model) to process input data and eventually generate a vulnerability prediction. In the prediction pipeline, code is input and processed into a format suitable for the predictive classifier. An attacker utilizing \tech{} provides code as well, but manipulates the code by utilizing SPTs to alter the feature space (e.g., graphs or embeddings used to represent code). These alterations in feature space are performed with the goal of avoiding a positive prediction by the target classifier. This approach preserves the functionality of the malicious code, since SPTs do not alter code functionality and maintain the same data type. However, it leads to a dangerous misclassification of vulnerable code as secure code. To aid understanding, we provide a more detailed mathematical formulation of our attack in Appendix \ref{appendix:mathematical-formulation}. The key points of our attack formulation are twofold. Firstly, we outline a realistic attack vector, user-provided code. Secondly, we note how minor manipulations (i.e., SPT application) in this initial input can propagate downstream in the prediction pipeline to induce prediction failure.

\begin{table*}[h]
\begin{center}
\vspace{-.1in}
\caption{Rules of natural transformations}
\vspace{-.1in}
\label{tab:transformation_summary}
\footnotesize
\begin{tabular}{c|c|c}
\toprule
{\bf Transformation} & {\bf Pre-transformation code example} & {\bf Post-transformation code example} \\
\midrule
{Assignment splitting} & {\example{1}} & {\example{2}} \\
\hline
{Compound assignment splitting} & {\example{3}} & {\example{4}} \\
\hline
{\texttt{\small while}-to-\texttt{\small for} conversion} & {\example{5}} & {\example{6}} \\
\hline
{\texttt{\small for}-to-\texttt{\small while} conversion} & {\example{7}} & {\example{8}} \\
\hline
{Condition negation} & {\example{9}} & {\example{10}} \\
\hline
{Condition splitting} & {\example{11}} & {\example{12}} \\
\hline
{Condition reordering} & {\example{13}} & {\example{14}} \\
\bottomrule
\end{tabular}
\end{center}
\vspace{-.15in}
\end{table*}

%% file: 4_transformations.tex
\section{Semantics Preserving Code Transformations}\label{sec:transformations}

\subsection{Designing Transformation Rules}

When designing transformation rules, our goal is to induce structural changes instead of single-element replacements that are employed in the relatively common identifier replacement transformation~\cite{springer2020strata, rabin2021generalizability, henkel2022semantic, zhang2020generating, yang2022natural, wang2023recode}. We propose changes that can affect the graph structure of the representation of targeted code (i.e., the node and edge in control and data dependency graphs). This specific choice of transformation rules aligns with our work's focus on graph-based vulnerability detectors. We justify this in Section \ref{subsec:effectiveness} by showing that \tech{}'s approach is more likely to affect the underlying graph representation of code than the existing work. Our specific rationale behind each chosen transformation is presented in Appendix \ref{appendix:transformation-choice}. 

We design the following types of transformations that can alter code in ways that an attacker desires without changing code semantics. With these rules, our work specifically explores the effect of SPTs which are more likely to affect graph-based models for VD.

\begin{itemize}[leftmargin=*]
    \item \textbf{Assignment splitting}: Given that an integer-valued expression of sufficient complexity is performed in the context of an assignment, the assignment can be split into multiple statements that perform the same operation in a step-by-step manner.
    \item \textbf{Compound assignment splitting}: In this case, specific compound assignment operators in our target languages can be split into an assignment and operator on the same line. 
    \item \textbf{Conversion between \texttt{\small while} and \texttt{\small for}}: Conversions between semantically equivalent \texttt{\small while} and \texttt{\small for} statements, which are performed by modifying statement components (i.e., loop conditions, update conditions, and initialization statements) as appropriate.
    \item \textbf{Condition negation}: Negation of the condition in an \texttt{\small if} statement to generate an equivalent \texttt{\small if} statement by rearranging the relevant code blocks. 
    \item \textbf{Condition splitting}: Rearranging \texttt{\small if} statements containing the ``and'' and ``or'' operations to generate \texttt{\small if} statements of equivalent effect by splitting and rearranging the pertinent code blocks. 
    \item \textbf{Condition reordering}: Swapping of elements in an \texttt{\small if} statement's condition. This kind of reordering can be performed given the absence of side-effecting expressions (such as assignments or function calls) in the condition.
\end{itemize}
Given the source code of a vulnerable function, there may be multiple sites where the same rule can be applied. It is also possible that multiple transformation rules can be performed. Therefore, we further define the following two kinds of transformations that
allow multiple locations and transformation rules to be combined:
\begin{itemize} [leftmargin=*]
    \item \textbf{Multi-location}: Application of a single prior transformation multiple times in the same source sample.
    \item \textbf{Multi-rule}: Application of multiple of any of the prior transformation rules at any location.
\end{itemize}
We perform relevant AST-based extractions as necessary to ensure that the relevant components for each SPT are known so that the transformation can be performed correctly. We also err on the side of caution to apply SPTs at sites in target code that are least likely to lead to semantic changes. Examples of pre- and post-application code for each of the initial set of SPTs can be found in \autoref{tab:transformation_summary}.

\subsection{Applying Transformations}\label{subsec:apply-transform}

To apply natural transformations to vulnerable code for generating adversarial examples, we perform the following steps: transformation site localization, rule selection, and transformation application. The procedure followed by \tech{} to carry out these steps is outlined in Algorithm \ref{algo:natgvd_generation}. We illustrate these steps using one sample from the Devign dataset~\cite{zhou2019devign}, which is vulnerable given an incorrect value of input parameter \texttt{\small y}, as shown in \autoref{lst:vulnerable_code}. 

\noindent\textbf{Transformation site localization.}
To apply an SPT to source code, the location where the transformation will be applied must first be determined. In this work, we locate SPT sites by identifying relevant locations in the target code's AST where transformations can be performed without violating semantic equivalency constraints. This process is implemented throughout \tech{} for C/C++ code with special checks to prevent transformations from being applied at unsafe locations where semantic changes might occur. To ensure the validity of generated adversarial code, \tech{} first avoids generating code from input data that cannot be parsed into a valid AST by LLVM (Clang). If a code sample fails to parse correctly, it will not pass the \textit{\textsc{ParseValid}} check in line \texttt{\small 3} of Algorithm \ref{algo:natgvd_generation}. For a valid sample of input code, \tech{} initially localizes to every possible node of the parsed AST. Then, it filters the targeted AST nodes such that no transformation will be applied where it would violate the semantics-preserving constraint. 
Example transformation sites located in this phase of \tech{} are indicated by the highlighted expressions in \autoref{lst:vulnerable_code}, which are found by searching for local patterns in the AST. \ref{algo:natgvd_generation}. As indicated in line \texttt{\small 6} of Algorithm \ref{algo:natgvd_generation}, location validity is determined in context of the transformation being applied. Validity constraints 
refer to whether a transformation would result in compilable and semantics-preserving code if performed at the given AST node. For example, if 
\texttt{\small f() < x} is being considered with the condition reordering transformation, it is not safe to reorder the condition, as \texttt{\small f()} may indirectly modify \texttt{\small x}. Therefore, given that input, $\textsc{\textit{ConstraintsValid}}(v, \mathcal{R}[t])$ returns \textsc{\small false} and the localization process would continue on to the next potential location instead.

\begin{lstlisting}[caption={Vulnerable code from Devign dataset}, label={lst:vulnerable_code}, belowskip=-0.1in, style=mystyle]
static inline void render(*@\_@*)line(*@\_@*)unrolled(intptr(*@\_@*)t x, intptr(*@\_@*)t y,
	int x1, intptr(*@\_@*)t sy, int ady, int adx, float *buf)
{
    int err = -adx;
    x -= x1 - 1;
    buf += x1 - 1;
    while (++x < 0) {
        err += ady;
        if ((*@\hlcyan{err >= 0}@*)) {
            err += ady - adx;
            y   += sy;
            buf[x++] = (*@ff\_vorbis\_floor1\_inverse\_db\_table[y];@*)
        }
        buf[x] = (*@ff\_vorbis\_floor1\_inverse\_db\_table[y];@*)
    }
    if ((*@\hlcyan{x <= 0}@*)) {
        if ((*@\hlcyan{err + ady >= 0}@*))
            y += sy;
        buf[x] = (*@ff\_vorbis\_floor1\_inverse\_db\_table[y];@*)
    }
}
\end{lstlisting}

\noindent\textbf{Rule selection.} Once a valid transformation location has been identified, the SPTs that can be applied to that point are identified and applied based on our proposed rules in \autoref{tab:transformation_summary}. The transformation rules are selected based on the potential changes that can be made at an AST node. All other rules will not be processed at that AST node. As seen in Algorithm \ref{algo:natgvd_generation}, all valid SPTs that can be applied at a given AST node will be generated. However, transformation rules that are illogical at that AST node will not be considered for application. For example, the \texttt{\small while}-to-\texttt{\small for} transformation rule cannot be logically applied at an \texttt{\small if} statement, so it will get skipped (i.e., at line \texttt{\small 5}, \textsc{\textit{CandidateNodes}} will not include the illogical node). If multiple locations are identified for a single SPT, \tech{} will apply it at each location in a one-to-one manner, generating multiple output samples. For the highlighted transformation sites at lines 9, 16, and 17 that are 
identified in Listing \ref{lst:vulnerable_code}, the condition reordering transformation can be applied at each location. We provide an 
example of files generated via multiple valid applications of a transformation in our Appendix \ref{appendix:multi-site-application}.

\noindent\textbf{Transformation application.} To apply an SPT, relevant parts of the local AST are extracted and rearranged into code that is semantically equivalent to, but differs from, original code. For example, in the \texttt{\small if}-based condition negation transformation, the full existing \texttt{\small if} statement is replaced (based on its structure) with an equivalent \texttt{\small if} statement that has its condition negated and branches swapped. For the assignment splitting transformation, appropriate variable types for temporary variables are also reverse engineered to preserve semantic consistency. For the first transformation site and corresponding SPT previously identified in 
\autoref{lst:vulnerable_code}, transformation application would change \texttt{\small{err >= 0}} to \texttt{\small{0 <= err}}. 

\noindent\textbf{Multiple-location and multi-rule application. }
To apply the multi-location and multi-rule transformations to source code, \tech{} applies compositions of SPTs on input code. Multi-location transformation is achieved by iteratively applying the same SPT to source code in strictly increasing location order. This strategy ensures termination if an SPT can be infinitely applied to the same location in the 
code. To apply the multi-rule transformation, \tech{} repeatedly applies all SPTs to the outputs of transformation application, until the desired number of transformation applications is reached. This simultaneously generates all combinations of SPTs as appropriate for the multi-rule scenario. In the code of \autoref{lst:vulnerable_code}, the multi-location transformation would additionally change \texttt{\small{x <= 0}} to \texttt{\small{0 >= x}} in the second transformation site and \texttt{\small{err + ady >= 0}} to \texttt{\small{0 <= err + ady}} in the third transformation site. Some examples of multi-location and multi-rule applications of transformations are demonstrated in Appendices \ref{appendix:multi-location-application} and \ref{appendix:multi-transformation-application}. 


\vspace{-0.01in}
{\footnotesize
\begin{algorithm}[htbp]
  \caption{\tech{}'s Adversarial Example  Generation}
  \label{algo:natgvd_generation}
  \SetAlgoLined
  \SetAlgoNoLine
  \DontPrintSemicolon
  
  \KwInput{$d$: Code sample, $\mathcal{T}$: Set of possible transformation rules, $\mathcal{R}$: Validity constraints for each transformation}
  \KwOutput{$\mathcal{F}$: Set of transformed files}
  $\mathcal{F} = \emptyset$ \tcp*[r]{ Initialize transformed file set} 
  $\mathcal{A} = \textsc{AST}(d)$\;
  \tcp{Check if AST was parsed correctly}
  \If{$\textsc{ParseValid}(A)$} {
    \ForEach{transformation rule $t \in \mathcal{T}$}{
      \tcp{Get AST nodes where $t$ can be applied}
      \ForEach{node $v \in \textsc{CandidateNodes}(\mathcal{A}, t)$}{
        \tcp{Ensure that $t$ is semantics-preserving at $v$}
        \If{\textsc{ConstraintsValid}($v$, $\mathcal{R}[t]$)}{
          $d' = t(d, v)$ \tcp*[r]{Apply SPT at $v$ to generate file}
          $\mathcal{F} = \mathcal{F} \cup \{d'\}$ \tcp*[r]{Add file to output}
        }
      }
    }
  }
  \algorithmicreturn{ $\mathcal{F}$}

\end{algorithm}}

%% file: 5_evaluation.tex
\vspace{-0.05in}
\section{Evaluation}\label{sec:evaluation}

In this section, we conduct extensive experiments to evaluate \tech{} by answering the following research questions:

\begin{itemize}[leftmargin=*]
\item\textbf{RQ1: Effectiveness.} How can \tech{} attack the recent representative graph-based vulnerability detection approaches? 

\item\textbf{RQ2: Efficiency.} How efficient and minimal are the generated adversarial examples?

\item\textbf{RQ3: Overhead.} What is the computational cost associated with generating adversarial examples?

\item\textbf{RQ4: Resilience.} Is \tech{} still effective towards a set of defense techniques?

\item\textbf{RQ5: Usability.} Are transformation rules in \tech{} natural from the developer perspective?

\end{itemize}

\begin{table*}[t]
\vspace{-.1in}
\begin{center}
\caption{Target vulnerability detection models}
\vspace{-.1in}
\label{tab:target_vd_summary}
\footnotesize
\begin{tabular}{c|c c c c}
\toprule
{\bf Method} & {\bf Graph-based (Y/N)} & {\bf Graph Structure} & {\bf Model Type} & {\bf Dataset} \\
\midrule
{Devign~\cite{zhou2019devign}} & \textcolor{teal}{\cmark} & {ncsCPG} & {GGNN} & {Devign} \\
\hline
{ReVeal~\cite{chakraborty2020deep}} & \textcolor{teal}{\cmark} & {CPG} & {GGNN} & {Devign} \\
\hline
{ReGVD~\cite{nguyen2022regvd}} & \textcolor{teal}{\cmark} & {Custom Graph} & {Transformer+GCN/GGNN} & {Devign} \\
\hline
{UniXcoder~\cite{guo2022unixcoder}} & \textcolor{teal}{\cmark} & { AST } & {Transformer} & {Big-Vul} \\
\hline
{LineVul~\cite{fu2022linevul}} & \textcolor{purple}{\xmark} & { - } & {Transformer} & {Big-Vul} \\
\bottomrule
\end{tabular}
\end{center}
\end{table*}

\begin{table*}[htbp]
\begin{center}
\vspace{-.1in}
\caption{Overall attack effectiveness (evasion rate) of transformations}
\vspace{-.1in}
\label{tab:effectiveness_results}
\footnotesize
\begin{tabular}{c|c c c c c}
\toprule
    \textbf{Transformation} & \textbf{Devign} & \textbf{ReVeal} & \textbf{ReGVD} & \textbf{UniXcoder} & \textbf{LineVul} \\
\hline
    One rule (at single location) 
    & 17.73\% & 39.34\% & 27.71\% & \;\,7.30\% & 1.81\% \\

    Multi-rule (each at single location)
    & 31.37\% & 51.99\% & 34.52\% & 12.72\% & \textbf{2.10}\% \\

    One rule (at single and all locations) + multi-rule (each at single location) 
    & \textbf{33.33}\% & \textbf{53.04}\% & \textbf{36.02}\% & \textbf{13.48}\% & 2.07\% \\

\bottomrule
\end{tabular}
\end{center}
\vspace{-.15in}
\end{table*}

\subsection{Effectiveness (RQ1)}\label{subsec:effectiveness}

\noindent\textbf{Target models.} We evaluate the effectiveness of \tech{} on SOTA graph-based deep learning models for function-level VD:
Devign~\cite{zhou2019devign}, ReVeal~\cite{chakraborty2020deep}, ReGVD~\cite{nguyen2022regvd}, and UniXcoder~\cite{guo2022unixcoder}, as summarized in \autoref{tab:target_vd_summary}. 
Specifically, Devign performs code embedding on a natural code sequence Code Property Graph (ncsCPG) by concatenating label-encoded type information and word2vec~\cite{mikolov2013word2vec} output for node code. It then employs a Gated Graph Neural Network (GGNN)~\cite{li2016gated_ggnn} model with recurrent layers and a 1-dimensional convolutional layer to generate output predictions.
ReVeal performs feature extraction on a CPG~\cite{yamaguchi2014modeling_cpg} using one-hot encoding and word2vec. ReVeal has similar model structure to Devign, but removes the 1D convolutional layer and instead uses a multi-layer perceptron (MLP) and a triplet loss function to extract predictions.
ReGVD initializes node features using a graph-based transformer (i.e., GraphCodeBERT~\cite{guo2020graphcodebert}) and then applies a residual-style GCN model to generate output predictions. 
We also fine-tune a pre-trained language model named UniXcoder~\cite{guo2022unixcoder}, currently the top model on Microsoft's CodeXGLUE leaderboard~\cite{codexglue_leaderboard} for the defect detection task, to perform function-level VD following the design of previous work~\cite{ni2023function}.

Besides the above GNN-based models and graph-based pre-trained language model, we also evaluate our generated adversarial examples on one non-graph-based model, LineVul~\cite{fu2022linevul}, providing insights on using \tech{} to attack graph-unaware models. LineVul applies byte-pair encoding (BPE)~\cite{sennrich2016neural_bpe} and a pre-trained CodeBERT model~\cite{feng2020codebert} directly to source code to identify lines or functions that contain vulnerable code. Note that LineVul provides function and line-level detection and our work targets the former.

\noindent\textbf{Datasets.} For a fair evaluation, we implement our attack on two datasets originally used by the above target models. 
For Devign, ReVeal, and ReGVD, we generate adversarial examples based on Devign's~\cite{zhou2019devign} dataset, which contains 27K labeled function-level C language samples from the FFmpeg and QEMU repositories hosted on GitHub. To ensure consistency between the model evaluations, we use standardized subsets for training/test/validation 
as specified by CodeXGLUE~\cite{lu2021codexglue}. 
For UniXcoder and LineVul, we apply the same dataset in the original work, i.e., Big-Vul~\cite{fan2020bigvul}, a C/C++ vulnerability dataset which includes approximately 177K non-vulnerable and 10K vulnerable functions from 348 popular GitHub projects.


\noindent\textbf{Metric.} 
We primarily examine the true positive samples generated by the VD systems. These samples are known to be vulnerable, and have also been identified by the VD systems as vulnerable. Therefore, the performance of models on the subset of data in our evaluation has 100\% predictive accuracy before SPts have been applied. We measure decrease in performance from this point to evaluate the effectiveness of our attack. In our case, we assume the attacker can generate multiple adversarial examples based on the transformation rules for each true positive sample. The attacker would intend to induce misclassification of these vulnerable code samples as non-vulnerable (i.e., change from true positive to false negative).  
In our evaluation, we use {\textit{evasion rate}} to measure the percentage of true positive predictions that can be turned into false negative predictions via adversarial attack. 

\begin{table*}[htbp]
\begin{center}
\caption{Attack effectiveness (evasion rate) of single and multiple locations for each transformation rule}
\vspace{-.1in}
\label{tab:effectiveness_results_multi}
\footnotesize	
\begin{tabular}{c||c|c|c|c|c}
\toprule
    \textbf{Transformation} & \textbf{Devign} & \textbf{ReVeal} & \textbf{ReGVD} & \textbf{UniXcoder} & \textbf{LineVul} \\
\hline\hline
    Assignment splitting 
    & \textbf{2.74}\% & \textbf{28.09}\% & 1.37\% & 0\% & 0\% \\
\hline
    Multi-location assignment splitting
    & 0\% & 19.05\% & \textbf{8.11}\% & 0\% & 0\% \\
\hline
\hline
    Compound assignment splitting 
    & 3.18\% & \textbf{41.67}\% & \textbf{5.71}\% & \textbf{5.41}\% & 0\% \\
\hline
    Multi-location compound assignment splitting 
    & \textbf{4.35}\% & 40.96\% & 5.09\% & 0\% & 0\% \\
\hline
\hline
    while-to-for conversion 
    & \textbf{11.11}\% & \textbf{35.80}\% & \textbf{2.33}\% & 0\% & \textbf{1.64}\% \\
\hline
    Multi-location while-to-for conversion 
    & 0\% & 31.25\% & 0\% & 0\% & 0\% \\    
\hline
\hline
    for-to-while conversion 
    & \textbf{17.45}\% & \textbf{37.29}\% & \textbf{6.48}\% & 0\% & 0\% \\
\hline
    Multi-location for-to-while conversion
    & 10.61\% & 26.92\% & 4.84\% & 0\% & 0\% \\
\hline
\hline
    Condition negation 
    & \textbf{16.56}\% & \textbf{49.15}\% & \textbf{29.91}\% & \textbf{8.19}\% & \textbf{1.62}\% \\
\hline
    Multi-location condition negation 
    & 14.61\% & 37.06\% & 19.18\% & 6.43\% & 1.75\% \\
\hline
\hline
    Condition splitting (``and'')
    & 12.68\% & 49.46\% & \textbf{2.33}\% & 0\% & \textbf{1.79}\% \\
\hline
    Multi-location condition splitting (``and'')
    & \textbf{25.00}\% & \textbf{50.00}\% & 0\% & 0\% & 0\% \\
\hline
\hline
    Condition splitting (``or'')
    & \textbf{7.41}\% & \textbf{43.42}\% & \textbf{2.22}\% & 0\% & 0\% \\
\hline
    Multi-location condition splitting (``or'')
    & 0\% & 30.77\% & 0\% & 0\% & 0\% \\
\hline
\hline
    Condition reordering
    & 1.88\% & \textbf{46.63}\% & \textbf{4.96}\% & 2.70\% & 0\% \\
\hline
    Multi-location condition reordering
    & \textbf{2.89}\% & 40.28\% & 2.54\% & \textbf{3.64}\% & 0\% \\
\hline
\bottomrule
\end{tabular}
\end{center}
\end{table*}

\begin{table*}[htbp]
\vspace{-.1in}
\begin{center}
\caption{Changes in code complexity by different transformation methods}
\vspace{-.1in}
\label{tab:complexity_increase}
\footnotesize
\begin{tabular}{c|c c c c}
\toprule

    \textbf{Method} & \textbf{Average CPG Degree} & \textbf{Lines of Code} & \textbf{Halstead Volume~\cite{halstead1977elements}} & \textbf{Cyclomatic Complexity~\cite{mccabe1976complexity}} \\
\hline
    MHM~\cite{zhang2020generating}
    & 0.004\% & 3.086\% & 0.000\% & 0.000\%\\

    Greedy-Attack~\cite{yang2022natural}
    & 0.000\% & 3.001\% & 0.000\% & 0.000\%\\

    ALERT~\cite{yang2022natural}
    & 0.000\% & 3.521\% & 0.000\% & 0.000\%\\

    \tech{} (ours)
    & \textbf{10.922}\% & \textbf{100.116}\% & \textbf{118.298}\% & \textbf{133.436}\%\\

\bottomrule
\end{tabular}
\end{center}
\vspace{-.15in}
\end{table*}

\noindent\textbf{Results.} 
To evaluate the effectiveness of \tech{}, we generate adversarial examples based on our proposed rules to attack the five aforementioned models. For each SPT in \autoref{tab:transformation_summary}, we first apply it at a single location within the target vulnerable sample. Next, we apply multiple transformation rules, with each rule applied at a single location. Finally, we generate adversarial examples by taking each rule and applying it at multiple locations within a target vulnerable sample. \autoref{tab:effectiveness_results} presents the effectiveness of \tech{} with evasion rate under the above settings.

\subsubsection{Effectiveness, comparisons, graph effects, and validation} We now discuss various aspects of our evaluation.

\noindent\textbf{Effectiveness with single-rule transformations.}
If the vulnerable code has already been submitted to a version control system like GitHub, each future code change will be recorded in a commit. In this case, to transform the vulnerable code into an adversarial example, the attacker may prefer to make modifications as minor as possible to avoid detection by other maintainers. Therefore, we evaluate the evasion rate when each SPT from \autoref{tab:transformation_summary} is applied at only one location. As shown in the second row of \autoref{tab:effectiveness_results}, \tech{} achieves an evasion rate of 17.73\%-39.34\% on three GNN-based VD models and 7.3\% on the graph-aware pre-trained model, which indicates the effectiveness of \tech{}. Note that, unlike vision models, in which AE attacks can usually achieve over 90\% evasion rate, our results exhibit effectiveness and naturalness when compared with previous works on code analysis-related models. We explain this further in Section \ref{sec:discussion}. 

\noindent\textbf{Effectiveness with multi-rule and multi-location transformations.}
We evaluate if applying multiple SPTs (each at a single location) can help enhance the effectiveness of an attack on graph-based models. From the third row of \autoref{tab:effectiveness_results}, we find that the evasion rate is increased by 62.5\% on average compared with modifying at a single location, which means that, if allowed in practice, incorporating more SPTs can improve the success rate of \tech{}. 
Applying all possible combinations of transformations to a code sample 
may generate an intractably large number of adversarial example candidates. Hence, we choose to evaluate combinations of two transformation rules to reduce the size of the search space.

With this in mind, we further apply all pairs of applicable transformations at suitable locations in the given code. By combining all generated samples until now, as shown in the last row of \autoref{tab:effectiveness_results}, we see that 
enabling all transformations described in Section \ref{subsec:apply-transform} provides the highest evasion rate, 33.33\%-53.04\% on three GNN-based VD models and 13.48\% on graph-aware pre-trained models.

\noindent\textbf{Comparison among graph and non-graph-based models.}
%
%
Analyzing the results in \autoref{tab:effectiveness_results}, we observe the high susceptibility of our tested GNN-based models (i.e., ReVeal, Devign, and ReGVD) to attacks based on natural SPTs. UniXcoder, a graph-aware transformer-based approach, also shows higher susceptibility than LineVul, which does not include code graph-based information during model pre-training.
Therefore, although graph-based deep learning models outperform 
other models, our results indicate they are vulnerable to adversarial examples generated by \tech{}. To account for this, developers should enhance these models' defense capabilities. However, there is a trade-off between VD performance and robustness. Although non-graph-based models generally perform worse than graph-based models, they may be more robust to SPTs.




\noindent\textbf{Effect on graphs.}\label{subsec:effect-on-graphs}
Because our proposed attack targets graph-based VD models, we analyze and quantify the graph-level differences before and after code modifications. We begin by statistically analyzing the changes in code complexity introduced by transformations from \tech{}. We use four metrics for this analysis: (1) {Average CPG Degree}, representing the average degree of a vertex in the Code Property Graph (CPG); (2) {Lines of Code}, tracking the number of lines in the code; (3) {Halstead Volume}~\cite{halstead1977elements}, which quantifies the information required to understand the code; and (4) {Cyclomatic Complexity}~\cite{mccabe1976complexity}, measuring the number of unique paths through the code. 
We compare our \tech{} with three other recent representative attack methods: (1) MHM~\cite{zhang2020generating}, which employs the Metropolis-Hastings algorithm to sample variables and replace them pseudo-randomly, using predicted class probabilities as guidance; (2) Greedy-Attack~\cite{yang2022natural}, which determines the importance of tokens for correct predictions and adversarially replaces them with natural tokens; and (3) ALERT~\cite{yang2022natural}, introduced in the same work as Greedy-Attack, which leverages a genetic algorithm-based strategy to optimize adversarial replacements with natural tokens.

As shown in Table \ref{tab:complexity_increase}, \tech{} leads to increases of 10.922\%, 100.116\%, 119.298\%, and 133.436\% across the four metrics, respectively. In contrast, the highest increases observed for the other three attack methods are 0.004\%, 3.521\%, 0\%, and 0\%. These results reveal significant graph-level changes caused uniquely by the transformation methods of \tech{}, demonstrating its ability to achieve a higher attack success rate compared to existing methods. Also, in Appendix~\ref{appendix:graph-effect-of-transformation}, we present a detailed example 
depicting how a small code change can 
impact subgraphs and structure of a CPG. 



\noindent\textbf{Validation of the generated adversarial code.}
Although we aim to design our adversarial code generation as a method to generate valid and correct code through SPTs, a separate verification is needed to evaluate the validity of the generated code. To this end, we perform a two-fold validation. 
First, to check if the generated code is valid, we write a script that uses Clang to automatically examine if each generated code files are compilable. The results show that all transformed files generated from the Devign dataset can compile successfully.
Second, to verify the correctness of the generated code (i.e., the functionality of the code after transformation still remains unchanged), we randomly select 100 files covering each type
of SPT employed by NatGVD for manual
inspection. The first two authors of this paper (each with 5+ years of experience in software development) manually compare the pre- and post-transformation code. We confirm that the generated adversarial code is semantically equivalent to the original code.

\subsection{Efficiency (RQ2)}\label{subsec:efficiency}

To determine how efficient and minimal the generated adversarial examples are, we perform per-rule analysis on the \textit{size} and \textit{evasion rate} of the SPTs. We perform overall analysis on SPT \textit{applicability} by measuring the percentage of true positive samples which transformations can be applied to. The size of an SPT is measured using qualitative analysis of code changes during transformation. 
The evasion rates of single-location SPTs are broken down by SPT type in \autoref{tab:effectiveness_results_multi}. This breakdown strategy lets us analyze whether skilled attackers would apply a single SPT at one or multiple locations, as well as whether applying a single SPT or combining multiple SPTs is more effective.
\subsubsection{Size}\label{subsec:size} 
Performing qualitative inspection on the transformations, we note that multi-location and multi-rule transformations have the potential to involve the most separate locations in a sample. Of the single-location SPTs, compound assignment splitting and condition reordering are the most compact changes. 
On the other hand, assignment splitting and condition splitting are the least compact changes. Assignment splitting can create a large amount of new lines based on the complexity of the expression being split up. Condition splitting is the worst transformation on the compactness metric, as splitting a condition can necessitate large amounts of code duplication to stay semantics-preserving. We perform a quantitative evaluation on generated adversarial examples using Levenshtein distance~\cite{levenshtein1966binary} as a measure of transformation. This evaluation reveals that, in the average case, most single-location transformations can be performed within under 100 character edits. 
We also observe that the average adversarial example from the multi-location and multi-rule transformations costs 319\% more character edits and 65\% more edits, respectively.

\subsubsection{Evasion rate} 
Analyzing the contents of \autoref{tab:effectiveness_results_multi}, we note that the multi-location transformation appears to make generated examples significantly less potent for most SPTs. In over half of the SPT-model combinations with non-zero attack effectiveness, the attacking potential of an SPT is hampered by applying it at multiple locations. Moreover, we note that the evasion rate of functions attacked via multi-location transformations is significantly less than the rate of ones which can be attacked via the single-location SPTs. Therefore, 
a skilled adversary is more likely to apply an SPT at a single location. An adversary would perhaps only perform similar changes across a function to maintain a consistent code style, which could be more likely to evade human inspection.


On the surface level, \autoref{tab:effectiveness_results_multi} indicates that condition negation is the most effective SPT in terms of evasion rate: it is applicable across all five target models, and has consistently high effectiveness compared with other SPTs for each model. However, taking into account the results from Section \ref{subsec:size}, we analyze the efficiency of each SPT. Compound assignment splitting is the most efficient SPT, causing an average of 0.96\% increased evasion rate per character changed (IER/CC). Condition negation, however, is only the 5th most efficient SPT, with 0.21\% IER/CC. Other efficient transformations include condition reordering, multi-location compound assignment splitting, and \texttt{\small while}-\texttt{\small to}-\texttt{\small for} conversion, with 0.80\%, 0.22\%, and 0.21\% IER/CC, respectively. On the other hand, the least efficient single and multi-location transformations are the single and multi-location variants of the condition splitting SPT for ``or'' conditions, inducing only 0.02\%  and 0.001\% IER/CC, respectively. Using this information, adversaries may craft less detectable attack strategies that apply the most efficient transformations first.


\subsubsection{Applicability}
We observe that the set of SPTs we define can be applied to 30.8\%-62.82\% of code samples in the true positives for each model. Since our implementation can parse 81.14\%-85.01\% of true positives across target models, we deduce an average successful application rate of 62\% for parsable samples by LLVM. We note that an adversary will almost certainly be able to compile and parse their own vulnerable code, so our set of SPTs would be applicable in over half of vulnerable function scenarios. Moreover, the SPTs are not logically complex to add to existing code, as they usually target a small amount of code for modification. This means that SPTs are very accessible for adversaries targeting graph-based models. Therefore, defending against such attacks is of high importance.

\subsection{Overhead (RQ3)}\label{subsec:overhead}

We use LLVM (Clang 18.0.0) for source code parsing and AST generation. Then, we design Python scripts to analyze the generated ASTs for program transformation to generate our adversarial examples. Linux shell scripts are developed to automate our data generation and testing pipelines. In total, \tech{} is implemented in Python and shell scripts with 7,000 LoC. Our experiments are conducted on an Ubuntu 22.04 server with an Intel W5-2455X running at 3.5 GHz, 64 GB RAM, and an NVIDIA A6000 with 48 GB Memory. For the target VD models, we follow the original environment settings in their released artifacts. 

Generally, launching adversarial example attacks using \tech{} has two steps: 1) performing transformations on vulnerable code to generate new adversarial examples, and 2) using these examples in a target VD system to mislead the model for misclassification. Since the cost for the latter is primarily determined by the target system, the overhead of \tech{} in the attacker's control is mainly on generating adversarial examples. 
In our experiments on UniXcoder, we start with 178 vulnerable functions for transformation. Applying all single-rule transformations to these 178 samples takes 30 seconds and produced 1881 new samples. Applying the multi-rule and multi-location transformations to these samples takes 341 seconds and produces 26947 new samples. This average, 80 samples generated per second, shows that \tech{} has acceptable overhead. 

\subsection{Resilience (RQ4)}\label{subsec:resilience}


\subsubsection{Ensemble methods}\label{subsubsec:ensemble}



Ensemble methods have been established as a defense against adversarial perturbations for deep neural networks~\cite{strauss2017ensemble}. This defense is based on an observation: an attack leading one model to misclassify may not imply the same for other networks performing the same task. In this subsection, we test the effectiveness of ensemble methods against \tech{}. We apply an ensemble model composed of the three graph-based VD systems (i.e., Devign, ReVeal, and ReGVD) with a majority voting scheme where the final classification results are based on the agreement of the highest number of constituent classifiers. 

Here, we select 13,870 adversarial examples that are generated by \tech{} and can deceive Devign, ReVeal, or ReGVD. By combining the classification results from these three models, the evasion rate can be decreased to 16.88\%, considering the evasion rate of single model Devign, ReVeal, and ReGVD is 33.33\%, 53.04\%, and 36.02\%, respectively.
However, using ensemble models can introduce many false positive predictions (i.e., the ensemble model incorrectly classifies the non-vulnerable samples as vulnerable ones).
Compared to using ReGVD alone, the number of false positives increases by 55.9\% 
and the precision decreases by 6.6\% when applying the ensemble model. Therefore, there is a trade-off between defending against adversarial examples generated by \tech{} and the practicality of ensemble model. While using ensemble techniques may offer some defense against adversarial examples, it can also generate too many false alerts, which can be unacceptable in real-world applications.



\subsubsection{Adversarial training}

\begin{figure}[t]
\centering
\captionsetup{justification=centering}
\includegraphics[width=0.65\linewidth, trim=1cm 0cm 1cm 0cm]
{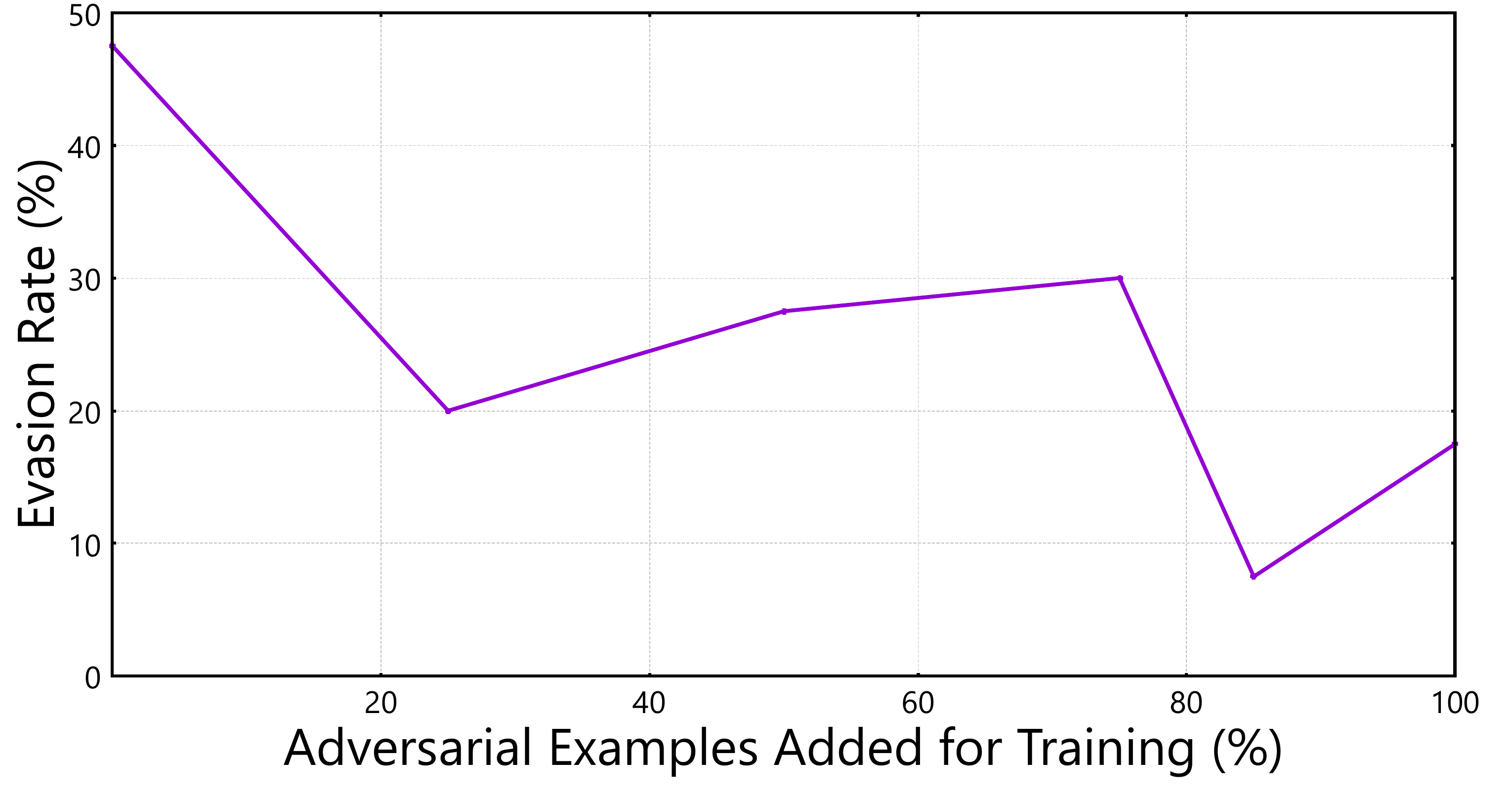}
\caption{Evasion rate after adversarial training on ReGVD.}
\label{fig:regvdtraining}
\end{figure}

Adversarial training is a technique that injects adversarial examples into model training data to increase robustness~\cite{bai2021recent}. We apply this 
in our setting by training a new ReGVD model with our generated adversarial examples. First, we divide 400 randomly selected true positive samples obtained from ReGVD in a 9:1 ratio. The second 10\% of samples are used as the test set. For the first 90\% of samples, we apply \tech{} to generate 24,835 adversarial examples and then gradually add parts of them into the original training set to train the ReGVD model.

The evasion rates with different ratios of adversarial examples added for training are shown in ~\autoref{fig:regvdtraining}. The base evasion rate (without including adversarial data in the training set) of ReGVD on the new test set is 47.5\%. The model becomes most effective in defending against \tech{} after introducing 85\% of the adversarial examples into the training set, with only 7.5\% evasion rate. We note 
that 
this augmented training set is almost double 
the size of the original dataset (i.e., from 21,854 to 42,963). It may be challenging to directly collect this many samples from the wild in practice. Also, such a large training set may incur significant preprocessing and training overhead, especially for deep learning-based VD systems. On the other hand, this also shows that adversarial training using ~\tech{} can enhance models' defense capabilities.

\subsection{Usability (RQ5)}\label{subsec:usability}

We conducted a user study to evaluate if the proposed transformations appear natural from a developer's perspective. We invited 10 participants, comprising 3 undergraduates, 1 master's student, 5 PhD students, and 1 postdoc, all with education backgrounds in computer science or software engineering. Among them, 1 participant has 5 years of software development experience, 3 have 5-10 years, and 6 have 10+ years of experience. In terms of their professional roles, 4 participants have worked as professional software developers, either in the past or currently. 6 participants are experienced security researchers. We presented the SPTs and examples shown in ~\autoref{tab:transformation_summary} to the participants and asked if these examples seemed natural to them, if they would write code in a similar style, and their thoughts on these coding styles.

All participants agreed that the SPTs were natural and would not find it strange to see code in similar styles in projects. For condition splitting and conversion from \texttt{\small while} to \texttt{\small for}, 4 out of 10 participants mentioned that they would not write code in these styles themselves but had seen others do so. One PhD student noted they would not personally use assignment splitting and compound assignment splitting but believed some beginners might prefer this style. For the other rules, all participants agreed they would write code in a similar style. 
Based on our user study, we believe the code generated by ~\tech{} is 
natural enough for experienced programmers to overlook abnormalities when reading ~\tech{}'s code in projects. We acknowledge that code naturalness can be subjective, but our proposed rules are relatively more natural compared to previous works which add dead and illegible code~\cite{li2023black_bagammo, imgrund2023broken}. 


%% file: 6_discussion.tex
\section{Discussion}\label{sec:discussion}
We look at similar attacks in the literature to understand the effectiveness of \tech{}. Yu et al. demonstrate an attack on Devign using a synthetic dataset with 65.20\% attack success rate~\cite{yu2023advulcode}.
Pour et al. evaluate an attack across multiple code-based tasks on code2seq~\cite{alon2018code2seq}, code2vec~\cite{alon2019code2vec}, and CodeBERT~\cite{feng2020codebert} with 5.41\% to 9.58\% effectiveness~\cite{pour2021search}. Applis et al. perform an attack against CodeBERT that causes a change in 22.6\%-27.6\% of comments generated~\cite{applis2021assessing}. Yang et al. ~\cite{yang2022natural} perform a natural attack with variable renaming on CodeBERT and GraphCodeBERT~\cite{guo2020graphcodebert}, achieving 7.96\%-48.79\% average attack success rates over three code analysis tasks. 
Our results of 13.483\%-53.04\% attack effectiveness indicate that existing graph-based VD methodologies are not sufficiently robust to \tech{}. Additionally, the examples generated by \tech{} are natural, which is not considered by most previous works. 

Normally, an adversarial attack needs to have some method to choose the attack rule and attack location with the best chance to generate a successful attack sample. For example, 
optimization techniques can be used to iteratively refine adversarial code until a successful attack sample is generated. However, we implement our attack under an extremely restricted variant of the black-box attack - 
we do not let our \tech{} see the prediction confidence of the targeted model, which makes a targeted attack impossible. 
This prevents the usage of works such as ALERT~\cite{yang2022natural} and MHM~\cite{zhang2020generating} which directly use prediction confidence to strengthen their attacks. 
If our scenario was less restrictive, 
then it would be trivial to adapt \tech{} to iteratively refine adversarial code. 
However, we show that targeted choice of adversarial transformations can increase attack performance even in the lowest information black-box scenario.



We focus on C/C++ at the source code level since these are the languages with the highest number of vulnerabilities and representative VD works, although \tech{} could theoretically be extended for code in different programming languages or at IR-level. Moreover, the idea of \tech{} can be transferable to other code analysis related tasks 
since graph-based models using code semantic structures also present superior performance on these code analysis-related tasks~\cite{guo2020graphcodebert,guo2022unixcoder,liu2024pre}. Adapting \tech{} to these tasks can help enhance the robustness of their models.

To the best of our knowledge, \tech{} is the first natural attack approach specifically focused on graph-based VD models. Existing works either require information that is not available in our scenario~\cite{zhang2022towards,yang2022natural}, fail to sufficiently modify graph data~\cite{yu2023advulcode, pour2021search, applis2021assessing}, overlook the crucial naturalness feature of practical adversarial examples~\cite{zhang2020generating}, or may struggle with datasets of extreme sizes~\cite{zhang2023black}. 
Unlike these earlier attacks, \tech{} is best aligned to attack graph-based models, exhibiting stable performance and requiring no 
unavailable information to generate natural adversarial examples. Moreover, the generated examples from \tech{}'s transformations can modify program graph representation through data flow, control flow, and data dependencies, which are not fully covered by other works.



%% file: 7_related_work.tex
\vspace{-.03in}
\section{Related Work}\label{sec:related_work}

\noindent\textbf{AE Attacks on LLM.} Works that perform adversarial attacks on Large Language Models for code-related tasks often include the variable renaming transformation with some success. Li et al.~\cite{li2022closer} explicitly single out this ``identifier transformation'' as a vulnerability of sequence- and AST-based transformer models and note the usage of AST data can help alleviate the issue. Zhang et al.~\cite{zhang2023black} extract existing variables in a codebase to rename variables, whereas Gao et al.~\cite{gao2023discrete} use common variable names from other codebases as substitutes. Yang et al.~\cite{yang2022natural} use sub-word token replacement suggestions from CodeBERT and GraphCodeBERT to do the same. 
Imgrund et al.~\cite{imgrund2023broken} replace all variable names with strings of twelve randomized lowercase letters. Applis et al.~\cite{applis2021assessing} do not specify a name generation strategy, but expand the scope of renaming to include variables, classes and methods. \tech{} forgoes variable renaming in favor of a clear focus on the graph nature of the target models.

\noindent\textbf{AE Attacks on GNN.}
Existing adversarial attacks on Graph Neural Networks often change graph structure/features or manipulate input data to deceive models. As an example of the former strategy, Yumlembam et al.~\cite{yumlembam2022iot} develop a methodology to add nodes and edges to generated adversarial Android API graphs. Li et al.~\cite{li2023black_bagammo} employ the latter strategy by inventing a ``try-catch trap'' that allows edges to be inserted in a function call graph (FCG). This innovation makes a graph-based attack possible through direct code modifications. For \tech{}, we adapt the ideas of a black-box scenario and affecting graphs through code modifications from the latter work.

Some works~\cite{yang2022natural, zhang2020generating, yefet2020adversarial, rabin2021generalizability, zhang2022towards} in the literature assess SPTs in the context of graph-based models. 
However, most works in this field only arbitrarily select
transformations and apply them to graph-based models. For example, Quiring et al.~\cite{quiring2019misleading} use SPTs such as integral type transformation and boolean transformation which are not expected to make large structural changes to graph-based code representations. \tech{} is the first work that has intentionally chosen SPTs with the sole purpose of affecting graph-based models.

Another major characteristic of the existing works~\cite{zhang2023black, li2022closer, zhou2022adversarial, wang2023recode, applis2021assessing, imgrund2023broken, srikant2021generating, rabin2021generalizability, quiring2019misleading, li2022semantic} is that they mainly target either combined (i.e., combining multiple individual SPTs) or individual transformations. We evaluate both types of transformation and additionally evaluate the effect of single-location and multi-location transformations, which is sparsely evaluated in the literature. These differences make our \tech{} well-rounded for evaluating the effects of SPTs against graph-based models as compared to other works.

\noindent\textbf{Naturalness in AE Attack.} 
Existing works that present AE attacks also raise the question of naturalness. Hendrycks et al.~\cite{hendrycks2021natural} raise the issue of adversarial examples that occur naturally in real-world scenarios. From another angle, Huang et al.~\cite{huang2023ala} present an AE attack that crafts natural-looking AEs for the image domain. Domains involving natural language and code also present this concern. For example, Wang et al.~\cite{wang2023recode} exclude function names, variable names, and type names from transformation outputs to ensure naturalness. Yang et al.~\cite{yang2022natural} propose the Naturalness Aware Attack (ALERT), which preserves naturalness even at the cost of decreased attack success rate. such as ~\cite{yang2022natural, hendrycks2021natural}, which emphasize the reduction of unnatural elements in AEs. 
\tech{} attempts to follow the goal of natural transformation in the context of graph-based models via SPTs which maintain a consistent code style over their scope or make small localized changes. 

%% file: 8_conclusion.tex
\section{Conclusion}\label{sec:conclusion}

We present \tech{}, a natural adversarial example attack on graph-based VD models including GNN and graph-aware pre-trained models. \tech{} applies a set of natural semantic-preserving transformations to generate adversarial vulnerable functions that can bypass target VD models. Evaluation on recent representative graph-based models shows the effectiveness of proposed attack.
Various existing defense techniques are also tested for its robustness. 

\section{Acknowledgment}\label{sec:acknowledgment}
This research is partially supported by the
National Science Foundation (NSF) grant CNS-2450602.

\clearpage

%% file: appendix_h.tex
\section{Mathematical formulation}\label{appendix:mathematical-formulation}
We outline in mathematical form the requirements of our attack, intermediate data transformations, and customizations for compatibility with an optimization objective and constraints. Although not strictly necessary to discuss our attack methodology, we believe such a formulation is helpful in understanding the general structure of \tech{}.

\noindent\textbf{Formulation for GNN-based models.} Most transformations in the attack occur in the context of a specific vulnerability detector $\mathpzc{D}$. As in work by Li et al.~\cite{li2023black_bagammo}, the input to the adversarial attack is source code which can be represented as a graph using a code-to-graph mapping $\mathpzc{M_{C\rightarrow G}^D}(\cdot)$. The graph is transformed into a vector for processing by the vulnerability detector's classifier using a graph-to-vector mapping $\mathpzc{M_{G\rightarrow V}^D(\cdot)}$. The classifier $\mathpzc{L^D(\cdot)}$ is an abstraction of the vector-to-prediction module of a GNN-based vulnerability detector. $\mathpzc{L^D}(\cdot)$ is used to produce a label which indicates whether the input is a vector representation of vulnerable or non-vulnerable code. The SPT $\mathpzc{T}$ operates on code $\mathpzc{c}$ and is validated using a functionality mapping $\mathpzc{M_F}$. $\mathpzc{M_F}$ is independent of $\mathpzc{D}$, taking inputs in the form of code and input data $\mathpzc{i}$ and mapping them to outcomes in terms of code outputs and side effects. Two pieces of code $\mathpzc{s}$ and $\mathpzc{t}$ are equivalent (but not necessarily identical) if $\mathpzc{M_F(s,\cdot)}\equiv\mathpzc{M_F(t,\cdot)}$, that is, $\mathpzc{\forall{i}\in{I(s)\cup{}I(t)},~M_F(s, i) = M_F(t, i)} $, where $\mathpzc{I(\cdot)}$ is the domain of possible input data for source code. Then, an attacker targeting GNN-based vulnerability detector $\mathpzc{D}$ needs SPT $\mathpzc{T}$ such that Equations \ref{eq:formulation} and \ref{eq:verify_formulation} hold for malicious code $\mathpzc{c}$:
\begin{equation} \label{eq:formulation}
    \mathpzc{L^D(M^D_{G\rightarrow V}(M^D_{C\rightarrow G}(c))) \neq L^D(M^D_{G\rightarrow V}(M^D_{C\rightarrow G}(T(c))))}
\end{equation}
\begin{equation} \label{eq:verify_formulation}
    \mathpzc{M_F(c,\cdot)}\equiv\mathpzc{M_F(T(c),\cdot)}
\end{equation}
Out of these elements, the attacker is aware of the initial source code $\mathpzc{c}$, the transformed source code $\mathpzc{T(c)}$, and whether the transformations are semantics-preserving (the truthiness of \autoref{eq:verify_formulation}). On the other hand, the implementation and outputs of the mappings $\mathpzc{M_{G\rightarrow V}^D(\cdot)}$ and $\mathpzc{M_{C\rightarrow G}^D}(\cdot)$ and the label $\mathpzc{L^D}$ are not available to the attacker due to the black-box setting. This means the attacker is required to create structural divergences between $\mathpzc{M^D_{C\rightarrow G}(c)}$ and $\mathpzc{M^D_{C\rightarrow G}(T(c))}$ which could induce changes in the outcomes of $\mathpzc{L^D}$ while still satisfying the constraints outlined through \autoref{eq:verify_formulation}.

Considering different ways to make targeted modifications, the structural divergences between $\mathpzc{M^D_{C\rightarrow G}(c)}$ and $\mathpzc{M^D_{C\rightarrow G}(T(c))}$ can be increased (by definition) if the graph representations of $\mathpzc{c}$ and $\mathpzc{T(c)}$ are significantly different. Changes that indirectly modify the graph-based representations of the code are most likely to induce classifier representation changes that will be reflected through $\mathpzc{M_{G\rightarrow V}^D(\cdot)}$. The attacker may prioritize maximized classifier inaccuracy with the consequence of extensive changes to these graphs. However, extensive code modifications can easily be noticed by others (for example, in scenarios where an attacker attempts to modify open source software to inject a vulnerability). Therefore, from the attacker's point of view, it would be beneficial to generate code transformations that efficiently change the AST, CFG, CDG, or DDG. 
In this work, simple changes (e.g., as outlined in Table \ref{tab:transformation_summary}) introduced via SPTs target graph-based VD systems. If these changes effectively induce classifier inaccuracy, it is reasonable that attackers might use these transformations as part of their attacks.

\noindent\textbf{Formulation for transformer-based models.} We note that models such as GraphCodeBERT and UniXcoder which are pre-trained with awareness of code structure of graphs do not use the graph-related code inputs to perform downstream tasks~\cite{guo2020graphcodebert, guo2022unixcoder}. 
Therefore, we adapt the GNN-based formulation for transformer-based vulnerability detectors and graph-based vulnerability detectors that do not compute an intermediate vector representation by removing the graph-related mappings $\mathpzc{M_{C\rightarrow G}^D}(\cdot)$ and $\mathpzc{M_{G\rightarrow V}^D}(\cdot)$ and replacing them with a unified mapping $\mathpzc{M_{C\rightarrow V}^D(\cdot)}$ which represents the code-to-vector processing performed by the transformer-based model. The attacker targeting transformer-based vulnerability detector $\mathpzc{D}$ needs SPT $\mathpzc{T}$ such that Equations \ref{eq:verify_formulation} and \ref{eq:formulation_llm} hold for malicious code $\mathpzc{c}$:
\begin{equation} \label{eq:formulation_llm}
    \mathpzc{L^D(M^D_{C\rightarrow V}(c)) \neq L^D(M^D_{C\rightarrow V}(T(c)))}
\end{equation}

This reformulation captures changes needed to apply consistent logic to GNN-based and transformer-based models. For example, $\mathpzc{L^D(\cdot)}$ is a valid abstraction of the vector-to-prediction module for 
these vulnerability detectors, and any methods of generating $\mathpzc{T(\cdot)}$ remain independent of mechanics of the model under attack.

\noindent\textbf{Formulation with optimization objective and constraints.}
The formulation we outlined earlier 
is modifiable to be compatible with further constraints and an optimization objective. These elements describe how further elements of an attack can be applied to derive an optimal SPT or set of SPTs. We mainly utilize constraint elements from this formulation in our work, leaving exploration of complex optimization objectives to future work. These constraints 
correspond to ensuring a transformed file is valid and has the same semantics as the pre-transformation file. 
Given a dataset $\mathpzc{X}$ of size $N$ consisting of various code samples $x_1 \cdots{} x_N$ and for which a set of constraints $\mathpzc{R}$ must be obeyed, \tech{}'s 
optimization objective is the following: 
\begin{equation}\label{eq:optimization_objective}
\begin{aligned}
\max_{\mathpzc{T_i \in T}} \quad & \sum_{j=1}^{N}{\mathbbold{1}_{\mathpzc{T}_i, x_j}}\\
\textrm{s.t.} \quad & \mathpzc{R(T}_i, x_j) \quad \begin{aligned}\forall \mathpzc{T}_i \in \mathpzc{T}, \\ \forall x_j \in \mathpzc{X}\end{aligned}
\end{aligned}
\end{equation}
where the indicator variable $\mathbbold{1}_{\mathpzc{T}_i, \;x_j}$ is assigned the value 1 if the prediction for code $x_j$ changes after $T_i$ is applied and 0 otherwise. 

\autoref{eq:optimization_objective} outlines how an attack might maximize the number of mismatches induced in $\mathpzc{L^D}(\cdot)$. 
This sample optimization objective demonstrates how SPTs analyzed in a similar way to our work can be used to identify an optimal SPT ${\mathpzc{T}_i}^*$ that, when applied alone, produces the highest classifier inaccuracy as compared to all available transformations $\mathpzc{T}$. A more complex optimization methodology~\cite{zhang2020generating, quiring2019misleading} 
may instead dynamically select a subset of transformations from $\mathpzc{T}$ to apply to a single code sample, achieving better evasion results. These complex optimization methodologies can be described using similar 
attack formulations, using the same idea of an optimization objective 
and a set of restrictions both applied across transformations and 
data. 

%% file: appendix_g.tex
\newcommand{\yes}{{\color{teal}\cmark}}
\newcommand{\no}{{\color{purple}\xmark}}

\begin{table*}[htbp]
\centering
\caption{Rationale behind choice of transformations}
\label{tab:transformation_choice}
\vspace{-3mm}
\begin{tabular}{>{\bfseries}lcm{4cm} m{6cm}}
\toprule
Rule Type / Proposed rule description & Included? & Modification Effects & Rationale \\ 
\midrule
Assignment splitting & \yes & Mainly affects data flow & All lines of the split assignment remain live and required for execution. At most, the transformed code would be labeled as inefficient. \\
Compound assignment splitting & \yes & Mostly affects AST by modifying the condition & The difference is so minor that it is basically indistinguishable to a human reviewer \\
{\texttt{\small while}-to-\texttt{\small for} conversion} & \yes & Minor effects on data flow between occurrences of the loop variable & Usage of \texttt{\small for}-based loops versus \texttt{\small while}-based loops is a matter of preference, and seeing either would typically not raise major concerns \\
{\texttt{\small for}-to-\texttt{\small while} conversion} & \yes & Minor effects on data flow between occurrences of the loop variable & Usage of \texttt{\small while}-based loops versus \texttt{\small for}-based loops is a matter of preference, and seeing either would typically not raise major concerns \\
Condition negation & \yes & Mostly affects control flow by inverting order of the AST subtrees & Using the inverse of the original condition and swapping the branches is typically a mundane difference to a human reviewer, especially if they have not seen the original condition \\
Condition splitting & \yes & Affects data flow and control flow by duplicating a region of code and routing control flow between them & A reviewer concerned about code duplication may be concerned, so this is the riskiest of the code transformations. This would be a bigger issue for conditions with larger bodies, so filtering could be added to prevent these cases \\
Condition reordering & \yes & Mostly affects AST by inverting order of subtrees and modifying the condition & The difference is so minor that it is basically indistinguishable to a human reviewer \\
Wrapping lines in \texttt{\small if(true)} & \no{} & Affects control flow of the AST subtree(s) corresponding to the wrapped lines & Dead condition, can easily be detected and eliminated via static analysis. Complex variants of \texttt{\small if(true)} can avoid static analysis but easily raise flags to a human reviewer \\
Variable renaming & \no & Affects neither control flow nor data flow & Tends to cause issues for LLM-based vulnerability detectors but is much less relevant to graph-based systems which learn from structural flows and dependencies \\
Inserting dead code statements & \no & Can modify control flow and data flow & Violates the assumption that the defender has the capabilities to detect dead code. \\
Inserting zero-equivalent expressions & \no & Can modify \texttt{def}-\texttt{use} relationships a little (e.g., by inserting \texttt{\small +(x*0)}), but mainly affects the AST structure in the expression & Sticks out to human reviewers and may also lead to warnings from code linters with redundant expression detection capabilities \\
Argument renaming/insertion & \no & Affects neither data flow nor control flow unless the number of parameters is modified & Either equivalent to variable renaming or completely changes call semantics (with the added potential of breaking API usages elsewhere in the code). Mostly irrelevant for graph-based changes while potentially creating additional problems \\
Return value modifications & \no & Potentially affects data dependencies related to connections in the call graph & Samples are generally isolated functions with unclear type dependencies, making it difficult to determine whether return types are preserved after transformation. \\

\bottomrule
\end{tabular}
\end{table*}

\FloatBarrier

\section{Transformation choice rationale}\label{appendix:transformation-choice}
Table \ref{tab:transformation_choice} presents key aspects we considered in our inclusion or exclusion of transformations from \tech{}'s attack strategy. We cover all transformations included in \tech{} as well as popular attack strategies that were not chosen for inclusion.


%% file: appendix_a.tex
\section{Multi-site transformation application}\label{appendix:multi-site-application}
The following example demonstrates how a code file with multiple valid sites for transformation can be used to generate multiple transformed files via \tech{}. In Listing \ref{lst:example_multisite_1}, multiple target sites for a single transformation are highlighted. In Listings \ref{lst:example_multisite_2} and \ref{lst:example_multisite_3}, \tech{}'s output for the original code sample is shown, with indication of local modifications. These modifications highlight \tech{}'s ability to perform multi-site transformation, generating multiple transformed output files from a single source file.

\vspace{-0.1cm}
\begin{lstlisting}[caption={Original code sample}, label={lst:example_multisite_1}, style=mystyle]
static int64(*@\_@*)t scene(*@\_@*)sad16(FrameRateContext *s, const uint16(*@\_@*)t *p1, 
    int p1(*@\_@*)linesize, const uint16(*@\_@*)t* p2, int p2(*@\_@*)linesize, int height)
{
    int64(*@\_@*)t sad;
    int x, y;
    for (sad = y = 0; y < height; (*@\hlcyan{y += 8}@*)) {
        for (x = 0; x < p1(*@\_@*)linesize; (*@\hlcyan{x += 8}@*)) {
            sad += sad(*@\_@*)8x8(*@\_@*)16(p1 + y * p1(*@\_@*)linesize + x,
                              p1(*@\_@*)linesize,
                              p2 + y * p2(*@\_@*)linesize + x,
                              p2(*@\_@*)linesize);
        }
    }
    return sad;
}
\end{lstlisting}

\vspace{-0.6cm}
\begin{lstlisting}[caption={First transformed code sample}, label={lst:example_multisite_2}, style=mystyle]
static int64(*@\_@*)t scene(*@\_@*)sad16(FrameRateContext *s, const uint16(*@\_@*)t *p1, 
    int p1(*@\_@*)linesize, const uint16(*@\_@*)t* p2, int p2(*@\_@*)linesize, int height)
{
    int64(*@\_@*)t sad;
    int x, y;
    for (sad = y = 0; y < height; (*@\hlcyan{y = y + (8)}@*)) {
        for (x = 0; x < p1(*@\_@*)linesize; x += 8) {
            sad += sad(*@\_@*)8x8(*@\_@*)16(p1 + y * p1(*@\_@*)linesize + x,
                              p1(*@\_@*)linesize,
                              p2 + y * p2(*@\_@*)linesize + x,
                              p2(*@\_@*)linesize);
        }
    }
    return sad;
}
\end{lstlisting}

\vspace{-0.6cm}
\begin{lstlisting}[caption={Second transformed code sample}, label={lst:example_multisite_3}, style=mystyle]
static int64(*@\_@*)t scene(*@\_@*)sad16(FrameRateContext *s, const uint16(*@\_@*)t *p1, 
    int p1(*@\_@*)linesize, const uint16(*@\_@*)t* p2, int p2(*@\_@*)linesize, int height)
{
    int64(*@\_@*)t sad;
    int x, y;
    for (sad = y = 0; y < height; y += 8) {
        for (x = 0; x < p1(*@\_@*)linesize; (*@\hlcyan{x = x + (8)}@*)) {
            sad += sad(*@\_@*)8x8(*@\_@*)16(p1 + y * p1(*@\_@*)linesize + x,
                              p1(*@\_@*)linesize,
                              p2 + y * p2(*@\_@*)linesize + x,
                              p2(*@\_@*)linesize);
        }
    }
    return sad;
}
\end{lstlisting}

\vspace{-0.6cm}

%% file: appendix_b.tex
\section{Multi-location transformation application}\label{appendix:multi-location-application}
The following example demonstrates how a code file with multiple valid sites for the same transformation can be used to generate a single transformed file via \tech{}. In Listing \ref{lst:example_multilocation_1}, multiple target sites for a single transformation are highlighted. In Listing \ref{lst:example_multilocation_2}, \tech{}'s output for the original code sample is shown, with indication of local modifications. This output highlights \tech{}'s ability to perform multi-location transformation, generating more complex transformed files than can be achieved using single-location application of semantics-preserving transformations.

\vspace{-0.1cm}
\begin{lstlisting}[caption={Original code sample}, label={lst:example_multilocation_1}, style=mystyle]
static int64(*@\_@*)t scene(*@\_@*)sad16(FrameRateContext *s, const uint16(*@\_@*)t *p1, 
    int p1(*@\_@*)linesize, const uint16(*@\_@*)t* p2, int p2(*@\_@*)linesize, int height)
{
    int64(*@\_@*)t sad;
    int x, y;
    for (sad = y = 0; y < height; (*@\hlcyan{y += 8}@*)) {
        for (x = 0; x < p1(*@\_@*)linesize; (*@\hlcyan{x += 8}@*)) {
            sad += sad(*@\_@*)8x8(*@\_@*)16(p1 + y * p1(*@\_@*)linesize + x,
                              p1(*@\_@*)linesize,
                              p2 + y * p2(*@\_@*)linesize + x,
                              p2(*@\_@*)linesize);
        }
    }
    return sad;
}
\end{lstlisting}

\vspace{-0.1cm}
\begin{lstlisting}[caption={Transformed code sample}, label={lst:example_multilocation_2}, style=mystyle]
static int64(*@\_@*)t scene(*@\_@*)sad16(FrameRateContext *s, const uint16(*@\_@*)t *p1, 
    int p1(*@\_@*)linesize, const uint16(*@\_@*)t* p2, int p2(*@\_@*)linesize, int height)
{
    int64(*@\_@*)t sad;
    int x, y;
    for (sad = y = 0; y < height; (*@\hlcyan{y = y + (8)}@*)) {
        for (x = 0; x < p1(*@\_@*)linesize; (*@\hlcyan{x = x + (8)}@*)) {
            sad += sad(*@\_@*)8x8(*@\_@*)16(p1 + y * p1(*@\_@*)linesize + x,
                              p1(*@\_@*)linesize,
                              p2 + y * p2(*@\_@*)linesize + x,
                              p2(*@\_@*)linesize);
        }
    }
    return sad;
}
\end{lstlisting}

%% file: appendix_c.tex
\section{Multi-transformation transformation application}\label{appendix:multi-transformation-application}
The following example demonstrates how a code file with multiple valid sites for different transformations can be used to generate transformed files via \tech{}. In Listing \ref{lst:example_multitransformation_1}, multiple target sites for multiple transformations are highlighted. In Listings \ref{lst:example_multitransformation_2}, \ref{lst:example_multitransformation_3}, and \ref{lst:example_multitransformation_4}, illustrative outputs for the original code sample are shown. These examples highlight \tech{}'s ability to perform multiple transformations at once, generating more diverse transformed files than can be achieved using single types of semantics-preserving transformations alone.

\vspace{0.0cm}
\begin{lstlisting}[caption={Original code sample}, label={lst:example_multitransformation_1}, style=mystyle]
static int64(*@\_@*)t scene(*@\_@*)sad16(FrameRateContext *s, const uint16(*@\_@*)t *p1, 
    int p1(*@\_@*)linesize, const uint16(*@\_@*)t* p2, int p2(*@\_@*)linesize, int height)
{
    int64(*@\_@*)t sad;
    int x, y;
    for (sad = y = 0; y < height; (*@\hlgreen{y += 8}@*)) {
        (*@\hlcyan{for (x = 0; x < p1\_linesize; }@*)(*@\hlred{x += 8}\hlcyan{) \{}@*)
            sad += sad(*@\_@*)8x8(*@\_@*)16(p1 + y * p1(*@\_@*)linesize + x,
                              p1(*@\_@*)linesize,
                              p2 + y * p2(*@\_@*)linesize + x,
                              p2(*@\_@*)linesize);
        (*@\hlcyan{\}}@*)
    }
    return sad;
}
\end{lstlisting}

\vspace{-0.3cm}
\begin{lstlisting}[caption={First transformed code sample (for-to-while conversion and compound assignment splitting)}, label={lst:example_multitransformation_2}, style=mystyle]
static int64(*@\_@*)t scene(*@\_@*)sad16(FrameRateContext *s, const uint16(*@\_@*)t *p1, 
    int p1(*@\_@*)linesize, const uint16(*@\_@*)t* p2, int p2(*@\_@*)linesize, int height)
{
    int64(*@\_@*)t sad;
    int x, y;
    for (sad = y = 0; y < height; y += 8) {
        (*@\hlcyan{x = 0;}@*)
        (*@\hlcyan{while(x < p1\_linesize)}@*)
        (*@\hlcyan{\{}@*)
            (*@\hlcyan{\{}@*)
                sad += sad(*@\_@*)8x8(*@\_@*)16(p1 + y * p1(*@\_@*)linesize + x,
                                  p1(*@\_@*)linesize,
                                  p2 + y * p2(*@\_@*)linesize + x,
                                  p2(*@\_@*)linesize);
            (*@\hlcyan{\}}@*)
            (*@\hlred{x = x + (8)}@*)(*@\hlcyan{;}@*)
        (*@\hlcyan{\}}@*)
    }
    return sad;
}
\end{lstlisting}

\vspace{-0.3cm}
\begin{lstlisting}[caption={Second transformed code sample (compound assignment splitting and for-to-while conversion)}, label={lst:example_multitransformation_3}, style=mystyle]
static int64(*@\_@*)t scene(*@\_@*)sad16(FrameRateContext *s, const uint16(*@\_@*)t *p1, 
    int p1(*@\_@*)linesize, const uint16(*@\_@*)t* p2, int p2(*@\_@*)linesize, int height)
{
    int64(*@\_@*)t sad;
    int x, y;
    for (sad = y = 0; y < height; (*@\hlgreen{y = y + (8)}@*)) {
        (*@\hlcyan{x = 0;}@*)
        (*@\hlcyan{while(x < p1\_linesize)}@*)
        (*@\hlcyan{\{}@*)
            (*@\hlcyan{\{}@*)
                sad += sad(*@\_@*)8x8(*@\_@*)16(p1 + y * p1(*@\_@*)linesize + x,
                                  p1(*@\_@*)linesize,
                                  p2 + y * p2(*@\_@*)linesize + x,
                                  p2(*@\_@*)linesize);
            (*@\hlcyan{\}}@*)
            (*@\hlcyan{x += 8;}@*)
        (*@\hlcyan{\}}@*)
    }
    return sad;
}
\end{lstlisting}

\vspace{-0.3cm}
\begin{lstlisting}[caption={Third transformed code sample (for-to-while conversion and while-to-for conversion)}, label={lst:example_multitransformation_4}, style=mystyle]
static int64(*@\_@*)t scene(*@\_@*)sad16(FrameRateContext *s, const uint16(*@\_@*)t *p1, 
    int p1(*@\_@*)linesize, const uint16(*@\_@*)t* p2, int p2(*@\_@*)linesize, int height)
{
    int64(*@\_@*)t sad;
    int x, y;
    for (sad = y = 0; y < height; y += 8) {
        (*@\hlcyan{x = 0;}@*)
        (*@\hlcyan{for(;x < p1\_linesize;)}@*)
        (*@\hlcyan{\{}@*)
            (*@\hlcyan{\{}@*)
                sad += sad(*@\_@*)8x8(*@\_@*)16(p1 + y * p1(*@\_@*)linesize + x,
                                  p1(*@\_@*)linesize,
                                  p2 + y * p2(*@\_@*)linesize + x,
                                  p2(*@\_@*)linesize);
            (*@\hlcyan{\}}@*)
            (*@\hlcyan{x += 8;}@*)
        (*@\hlcyan{\}}@*)
    }
    return sad;
}
\end{lstlisting}

\vspace{0.3cm}

%% file: appendix_d.tex
\section{Graph effect of natural transformations}\label{appendix:graph-effect-of-transformation}
\begin{figure}[H]
\centering
\captionsetup{justification=centering}
\includegraphics[width=1\linewidth, trim=1cm 1cm 1cm 1cm]{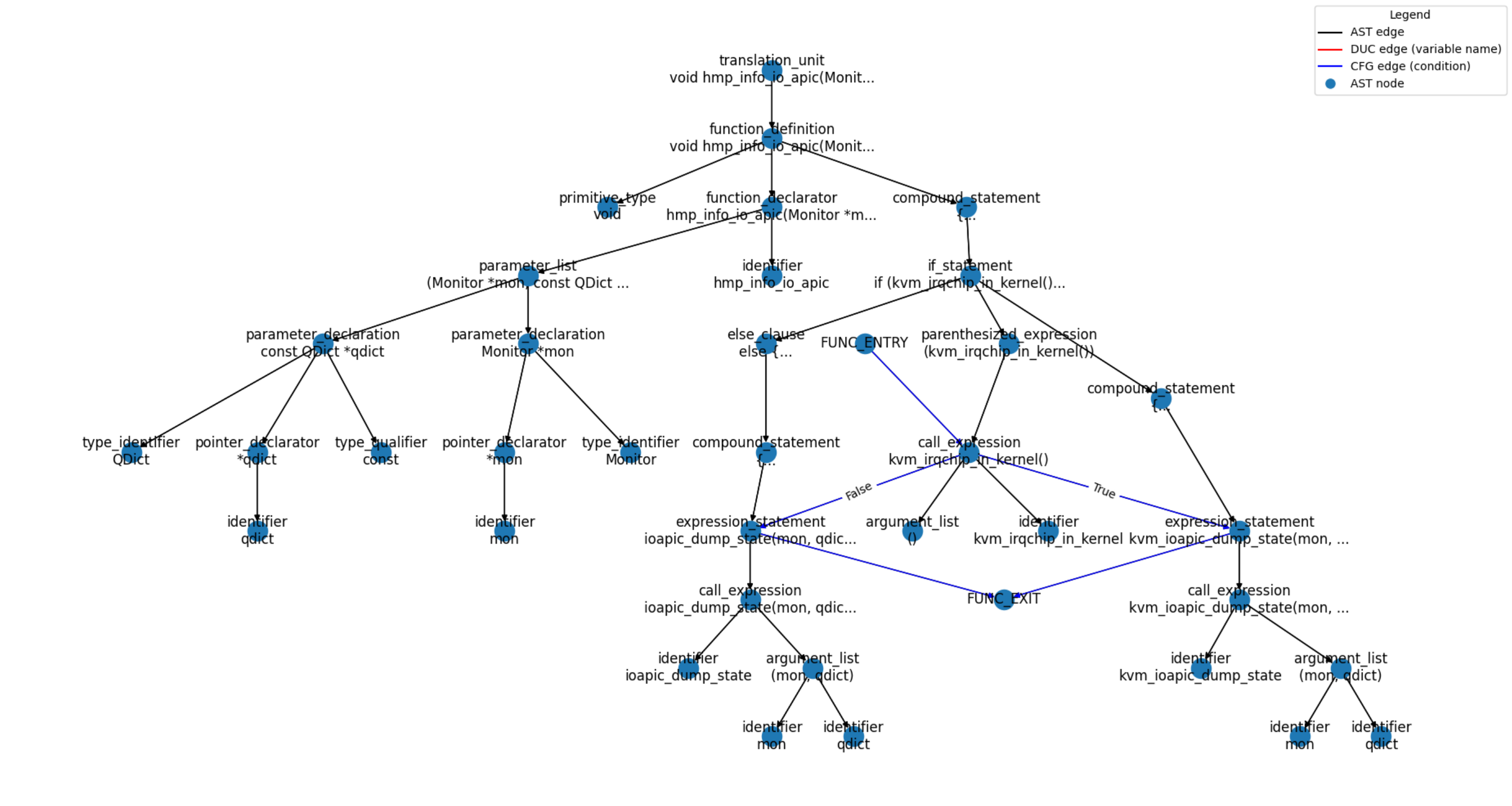}
\caption{CPG for original code}
\Description{Code Property Graph for original code}
\label{fig:cpg-orig-graph-effect}
\end{figure}

\begin{figure}[H]
\centering
\captionsetup{justification=centering}
\includegraphics[width=1\linewidth, trim=1cm 1cm 1cm 1cm]{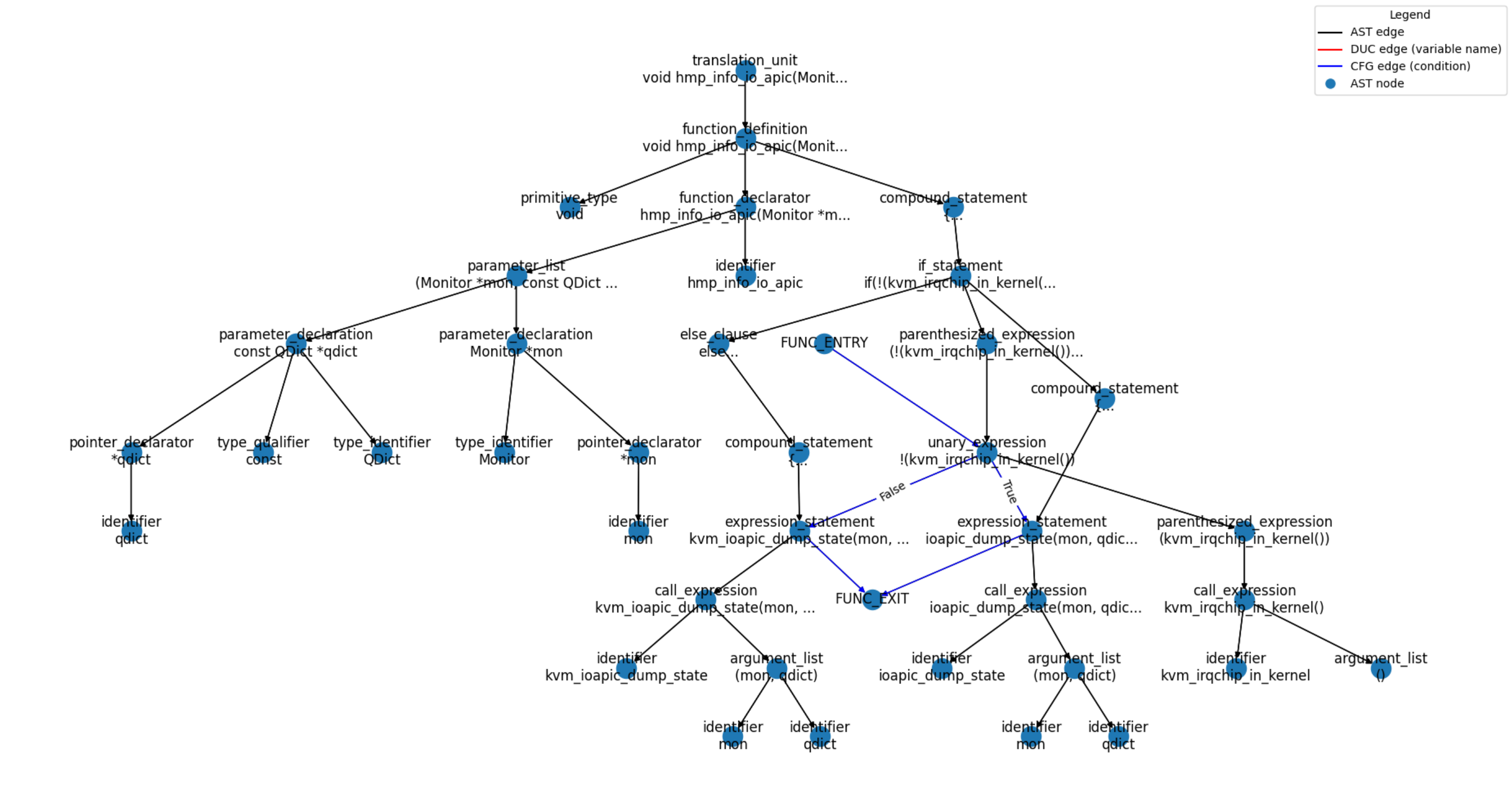}
\caption{CPG for transformed code}
\Description{Code Property Graph for transformed code}
\label{fig:cpg-transf-graph-effect}
\end{figure}

We now present an example of the changes induced through the application of a natural semantics-preserving transformation and observe the effects on a custom CPG generated by the tree-climber tool\footnote{https://github.com/bstee615/tree-climber}. Unlike the classic definition of CPG, the CPGs used for visualization purposes consist of AST nodes which are connected by AST edges, Def-Use Chain (DUC) edges, and CFG edges. We first present original code in Listing \ref{lst:example_graph_effect_1} and its corresponding CPG in Figure \ref{fig:cpg-orig-graph-effect}. We then present the code after the condition negation transformation has been applied in Listing \ref{lst:example_graph_effect_2}, with the corresponding CPG in Figure \ref{fig:cpg-transf-graph-effect}. Visual observation reveals that applying the transformation results in the creation of two new AST nodes and alters the structure of the graph by increasing graph traversal distance between certain nodes in the CPG. The creation of new AST nodes is related to the negation of the boolean condition in the \texttt{\small if} statement, and two subtrees in the graph which are connected at different CFG edges correspond to the switched branches of the \texttt{\small if} statement. These changes provide tangible evidence that \tech{} is able to induce changes in feature space (i.e., graphs used to represent code) through its transformation rules.

\vspace{-0.3cm}

\vspace{-0.3cm}
\begin{lstlisting}[caption={Original code}, label={lst:example_graph_effect_1}, style=mystyle]
void hmp(*@\_@*)info(*@\_@*)io(*@\_@*)apic(Monitor *mon, const QDict *qdict)
{
    if (kvm(*@\_@*)irqchip(*@\_@*)in(*@\_@*)kernel()) {
        kvm(*@\_@*)ioapic(*@\_@*)dump(*@\_@*)state(mon, qdict);
    } else {
        ioapic(*@\_@*)dump(*@\_@*)state(mon, qdict);
    }
}
\end{lstlisting}

\vspace{-0.3cm}
\begin{lstlisting}[caption={Transformed code}, label={lst:example_graph_effect_2}, style=mystyle]
void hmp(*@\_@*)info(*@\_@*)io(*@\_@*)apic(Monitor *mon, const QDict *qdict)
{
    if(!(kvm(*@\_@*)irqchip(*@\_@*)in(*@\_@*)kernel()))
    {
        ioapic(*@\_@*)dump(*@\_@*)state(mon, qdict);
    }
    else
    {
        kvm(*@\_@*)ioapic(*@\_@*)dump(*@\_@*)state(mon, qdict);
    }
}
\end{lstlisting}